\documentclass{vldb}

\usepackage{graphicx}
\usepackage[tight,footnotesize]{subfigure}
\usepackage{balance}  
\usepackage{algpseudocode}
\usepackage{algorithm}
\usepackage{mathtools}
\usepackage{amsmath}
\usepackage{amssymb}
\usepackage{subfigure}
\usepackage{ascmac}
\usepackage{multirow}
\usepackage{nidanfloat}

\newtheorem{proposition}{Proposition}
\newtheorem{definition}{Definition}
\newtheorem{example}{Example}
\newtheorem{theorem}{Theorem}
\newtheorem{lemma}{Lemma}

\newcommand{\algrule}[1][.2pt]{\par\vskip.5\baselineskip\hrule height #1\par\vskip.5\baselineskip}

\newdimen\tbaselineshift

\begin{document}


\title{Approximate-Closed-Itemset Mining for \\Streaming Data Under Resource Constraint
}



%
%
%
%

\numberofauthors{3} 

\author{
%
%
\alignauthor
Yoshitaka Yamamoto\\
       \affaddr{University of Yamanashi, Japan}\\
       \affaddr{yyamamoto@yamanashi.ac.jp}
\alignauthor
Yasuo Tabei\\
       \affaddr{RIKEN Center for Advanced Intelligence Project, Japan}\\
       \affaddr{yasuo.tabei@riken.jp}
\alignauthor
Koji Iwanuma\\
       \affaddr{University of Yamanashi, Japan}\\
       \affaddr{iwanuma@yamanashi.ac.jp}
}

\maketitle

\begin{abstract}
Here, we present a novel algorithm for frequent itemset mining 
for streaming data (FIM-SD).
%
For the past decade, various FIM-SD methods in one-pass approximation settings have been developed to approximate the frequency of each itemset.
These approaches can be categorized into two approximation types: {\it parameter-constrained} (PC) mining and {\it resource-constrained} (RC) mining.
PC methods control the maximum error that can be included in the frequency based on a pre-defined parameter. 
In contrast, RC methods limit the maximum memory consumption based on resource constraints.
However, the existing PC methods can exponentially increase the memory consumption, while the existing 
RC methods can rapidly increase the maximum error. 
%
In this study, we address this problem by introducing the notion of a condensed representation, called a {\it $\Delta$-covered set}, to the RC approximation. 
This notion is regarded as an extension of the closedness compression and when $\Delta = 0$, the solution corresponds to an ordinary closed itemset.
The algorithm searches for such approximate closed itemsets that can restore the frequent itemsets
and their frequencies under resource constraint while the maximum error is bounded by an integer, $\Delta$.
We first propose a one-pass approximation algorithm to find the condensed solution. 
Then, we improve the basic algorithm by introducing a unified PC-RC approximation approach.
%
Finally, we empirically demonstrate that the proposed algorithm significantly outperforms the state-of-the-art PC and RC methods for FIM-SD.
\end{abstract}


\section{Introduction}
{\em Streaming data analysis} is a central issue in many domains such as computer system monitoring~\cite{du16}, online text analysis~\cite{Iwata10,Hoffman10}, financial and economic analyses~\cite{Zhao13, Shu02}, and medical and health record data analysis~\cite{Ginsberg09, Keogh01}.
{\em Streaming data} is an infinite and continuous sequence of data, and, nowadays, is generated and collected rapidly. 
The sudden emergence of an intensive bursty event, called {\it concept drift}, in streaming data can make it difficult to extract meaningful information from such data. 
Therefore, there is a strong and growing need for powerful methods that are robust against concept drift in large-scale streaming data analysis. 

{\em Frequent itemset mining for streaming data (FIM-SD)}~\cite{han07} is the most fundamental and well-used task in streaming data analysis. 
It is used to find frequently occurring itemsets (in this study, sets of non-negative integers) in streaming data 
(in this study, a sequence of itemsets). 
FIM-SD must exhibit two important properties;
(i) the {\em real time property}, which is the ability to process a huge volume of itemsets continuously arriving at high speed and simultaneously 
(on-the-fly) output the detected frequent itemsets (FIs);
and (ii) {\em memory efficiency}, which is the ability to  enumerate FIs while managing an exponential number of candidate FIs with limited memory.
Concept drift suddenly introduces huge itemsets that must be processed, which demands a huge amount of memory. This makes it difficult to design scalable and efficient FIM-SD. Thus, the development of a space-efficient FIM-SD that is especially robust against concept drift is an important open challenge that must be solved to enable large-scale streaming data analysis.

Two major classes of solutions to compactly represent FIs
in FIM-SD have been proposed: {\em closed} FIs, which have no frequent supersets with the same support, and {\em maximal} FIs, which are not contained in any other FI.
Mining closed and maximal itemsets allows the number of solutions to be reduced by orders of magnitude compared with the mining of all FIs, which significantly reduces the memory consumed by FIM-SD. However, since an exponential number of closed or maximal itemsets may still exist, the risk of huge memory consumption by FIM-SD remains.

\begin{figure}[t]
\vspace{-22mm}
\begin{center}
  \includegraphics[width=8.5cm]{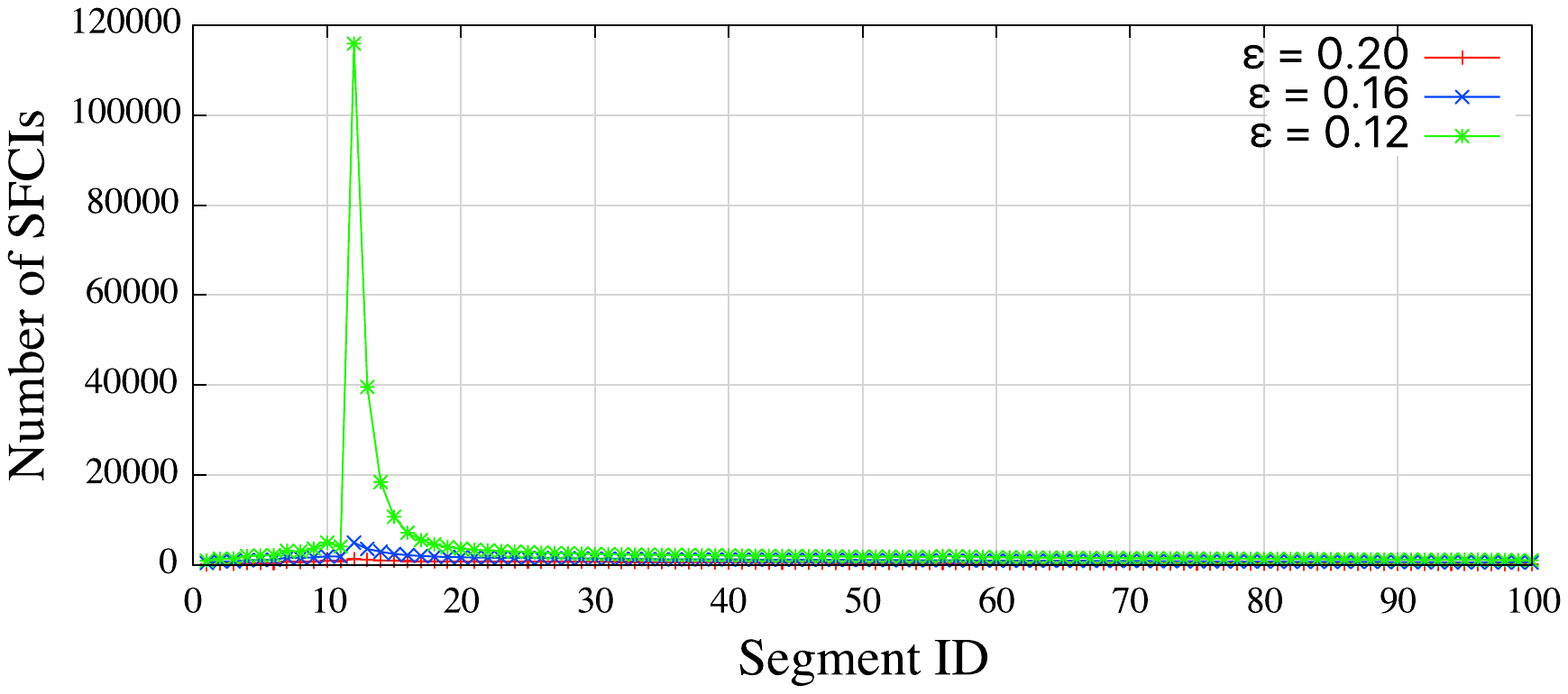}
  \caption{Time-series on the number of SFCIs}	
  \vspace{-7mm}	
  \label{moti_ex}
\end{center} 
\end{figure}

Recent research on FIM-SD~\cite{manku02,jin05,song07,li08,cheng08,shin14,yamamoto14,quadrana15} has mainly pursued  two directions:
{\em parameter constrained (PC) mining} and {\em resource constrained (RC) mining}. 
In PC mining~\cite{xin05,cheng06,cheng08,shin14,quadrana15}, an intermediate solution between closed and maximal itemsets, called a {\em semi-frequent closed itemset (SFCI)}, is computed; a parameter, $\epsilon$, controls 
the maximum error that can be included in the restored frequencies. 
Although PC mining enables a huge memory reduction 
compared with closed and maximal itemset mining, it still consumes huge amounts of memory when processing concept drift~\cite{hu17,boley10}.
%
%
%
This is illustrated in Figure~\ref{moti_ex}, which shows examples of the number of SFCIs 
mined by MOA-IncMine~\cite{quadrana15} which is the state-of-the-art PC method, in a dataset of a hadoop grid logs~\cite{yahoo}. 
The results show that the number of SFCIs bursts when processing the 12th segment of transactions 
(which represents concept drift) in the case when $\epsilon$ is low (0.12), resulting in huge memory consumption. 

On the other hand, RC mining is another approximation working under resource (i.e., memory) constraint; 
Skip LC-SS is a representative RC mining method~\cite{yamamoto14}.  
RC mining works in a constant space based on a user-defined parameter and returns approximate solutions of FIs;  
this represents a significant advantage. However, a major disadvantage of RC mining is the large error in approximate solutions due to the strict space constraint, rendering some solutions in the output useless~\cite{yamamoto14}. 

\smallskip
{\em Contributions.}
In this study, we tackle the problem of approximation error in RC mining by introducing a novel condensed representation called a {\em $\Delta$-covered set}. 
The notion of $\Delta$-cover is regarded as an extension of closedness compression and allows the original FIs to be compressed while bounding the maximum error by an integer, $\Delta$. 

First, we propose a new RC method to find this condensed solution. 
The approach involves two key techniques: {\it incremental intersection}, which is an incremental way to compute the closed itemsets~\cite{borgelt11,yen11}, and {\it minimum entry deletion}, which is a space-saving technique for the RC approximation~\cite{metwally05,yamamoto14}.
Consequently, the proposed method exhibits the following three characteristics:
\begin{description}
\item[On-the-fly manner:] it can process any transaction that has $L$ items in $O(k)$ space in almost $O(kL)$ time.
\item[Anytime feature:] it can monitor the maximum error, $\Delta$, and return the output, $T$, at anytime. 
\item[Quality of output:] for every FI $\alpha$, $T$ contains some superset $\beta$ of $\alpha$ such that the difference between the frequencies of $\alpha$ and $\beta$ is at most $\Delta$.
\end{description}
%


Next, we present a modification of this baseline method.
As shown in Figure~\ref{moti_ex}, sudden and intensive memory consumption can be considered a temporary phenomenon. 
Indeed, normal PC approximation is sufficient except for within a brief period. 
From the viewpoint of memory efficiency, it is reasonable to switch between PC and RC approximations. 
Thus, in this study, we call such a ``unified'' approach {\it PARASOL}. In PARASOL, the memory consumption is generally controlled using an error parameter and an RC approximation using a size constant is taken only in certain scenarios. 
In addition, we present a post-processing technique called {\it $\Delta$-compression} to reduce the size of the raw output to yield a more concise $\Delta$-covered set.
Moreover, we investigate a data structure to efficiently carry out the key operations in the proposed method.
Finally, we empirically show that the proposed method outperforms the existing PC and RC methods when applied to various real datasets.

The remainder of the paper is organized as follows.
Section~2 is a preliminary. 
Section~3 presents a brief review of the existing methods.
Section~4 describes the baseline algorithm used to compute a $\Delta$-covered set, and 
Section~5 describes PARASOL and $\Delta$-compression.
The utilized data structure is discussed in Section~6.
The experimental results are summarized in Section~7,
and we conclude in Section~8. 

%

\section{Notation and Terminology}
Let $I = \{1, 2, \ldots, N \}$ be the universal set of items. 
Itemset $t$ is a non-empty subset of $I$, i.e., $t \subset I$. 
Data stream ${\mathcal S}_n$ is the sequence of itemsets $\langle t_1, t_2, \ldots, t_n \rangle$ for which $t_i \subset I$ for $i=1,2,...,n$ where $n$ denotes the timestamp for which an output is requested by the user. 
Each $t_i$ is called transaction at timestamp~$i$.
The number of items in $t_i$ (i.e, the cardinality) is denoted as 
$|t_i|$ and is referred to as the {\it length} of $t_i$.
$L$ denotes the maximum length of the transactions in ${\mathcal S}_n$.
For itemset $\alpha\subset I$ and timestamp $i$, $tran(\alpha, i)$ denotes the family of itemsets that include $\alpha$ as a subset at timestamp $i$ (i.e., $tran(\alpha, i) = \{t_j~|~\alpha\subseteq t_j,\  1\leq j \leq i\})$.
Support $sup(\alpha, i)$ of itemset $\alpha$ at timestamp $i$ is defined as $|tran(\alpha, i)|$.
%
Given a minimum support threshold $\sigma$ ($0 \leq \sigma \leq 1)$,
if $sup(\alpha, n) > \sigma n$, then $\alpha$ is frequent with respect to $\sigma$ in ${\mathcal S}_n$.
${\cal F}_n$ denotes the family of FIs with respect to $\sigma$ at timestamp $n$ in ${\mathcal S}_n$, 
i.e., ${\cal F}_n = \{\alpha \subset I| \sup(\alpha, n) > \sigma n\}$.
FI $\alpha$ is {\it closed} if there is no $\alpha$'s proper superset whose frequency is equal to $\alpha$'s frequency in ${\cal F}_n$.
On the other hand, FI $\alpha$ is {\it maximal} if there is no $\alpha$'s proper superset in ${\cal F}_n$.

\subsection{Problem setting}
In this paper, to overcome the limitation of the state-of-the-art PC and RC methods, one key idea is to formalize a novel FIM-SD problem in which we seek a condensed representation of FIs with limited memory. 
This condensed representation is defined as follows:

\begin{definition}[$\Delta$-cover]
Let $\alpha$ and $\beta$ be two itemsets. 
If $\alpha \subseteq \beta$ and $sup(\alpha, i) \leq sup(\beta, i) + \Delta$ 
for a non-negative integer $\Delta$, then
$\alpha$ is {\em $\Delta$-covered} by $\beta$ at timestamp~$i$; 
this state is denoted by $\alpha \preceq_{\Delta}^i \beta$.
\end{definition}
\begin{definition}[$\Delta$-covered set]
Let $P$ and $Q$ be two families of itemsets.
If $\forall \alpha\in P$ $\exists \beta \in Q$ such that  $\alpha \preceq_\Delta^i \beta$ for a non-negative integer $\Delta$, 
then $Q$ is a {\em $\Delta$-covered set} of $P$ at timestamp $i$. 
\end{definition}

\begin{figure}[h]
  \vspace{-5mm}	
	\begin{center}
		\includegraphics[width=7cm]{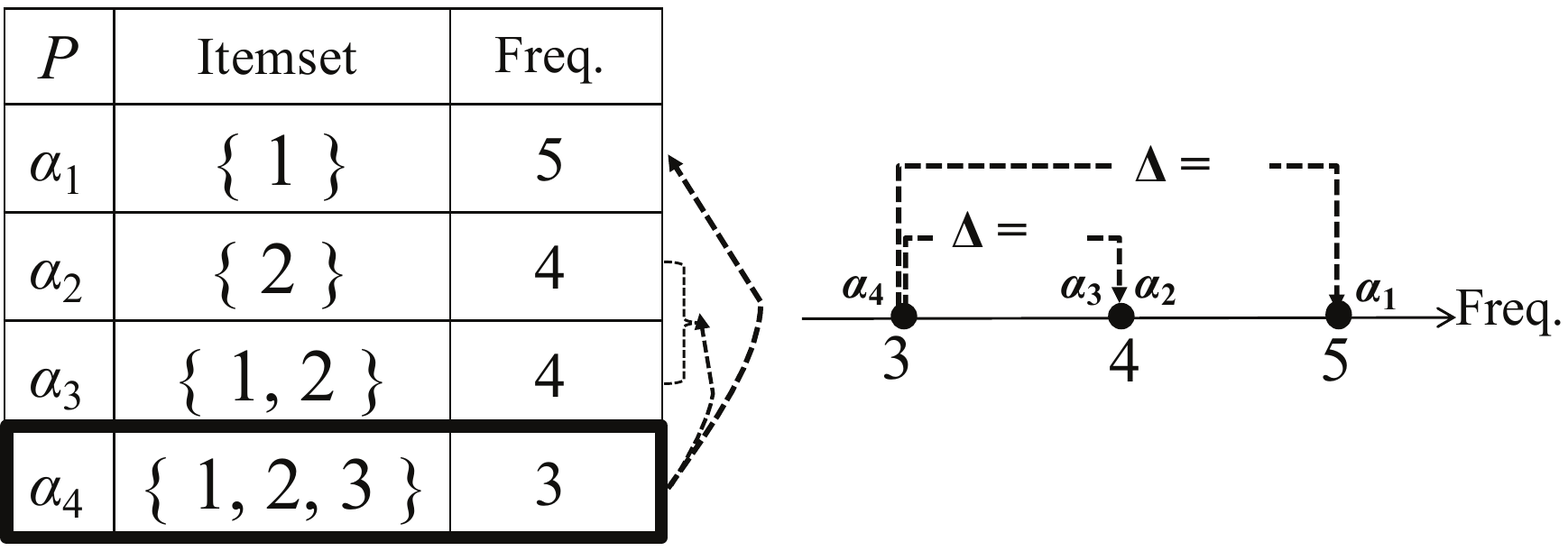}
		  \vspace{-5mm}	
		\caption{Example of a $\Delta$-covered set.}
                \label{fig:keyidea}
	\end{center}
  \vspace{-7mm}	
\end{figure}
\begin{example}\label{ex2}
Consider the family, $P$ of four itemsets $\alpha_1$, $\alpha_2$, $\alpha_3$, and $\alpha_4$, shown in Figure~\ref{fig:keyidea}.
For $Q_0 = \{\alpha_1, \alpha_3, \alpha_4\}$, $Q_1 = \{\alpha_3, \alpha_4\}$ and $Q_2 = \{\alpha_4\}$. 
Hence, $Q_0$, $Q_1$ and $Q_2$ are $0$-covered, $1$-covered and $2$-covered sets of $P$, respectively. 
\end{example}

A $\Delta$-covered set is a generalization of the closed itemsets.
A $0$-covered set of the family ${\cal F}$ of FIs is equal to the closed itemsets of the same family ${\cal F}$.
Furthermore, each FI and its frequency can be recovered from a 
$\Delta$-covered set with an error that is bounded by $\Delta$; 
this property is called {\em $\Delta$-deficiency}.
Let us consider the 2-covered set $Q_2$, in Example~\ref{ex2}.
Given anti-monotonicity with respect to frequency, $sup(\alpha_4, 5) \leq sup(\alpha_i, 5)$ ($1\leq i \leq 3$) holds.
Since $Q_2$ is a 2-covered set of $P$, $\alpha_4\preceq_{2}^5 \alpha_i$ so $sup(\alpha_i, 5)\leq sup(\alpha_4, 5) + 2$.
Since both $3\leq sup(\alpha_i, 5)$ and $sup(\alpha_i, 5) \leq 5$ hold, the error is bounded by $2$.

The objective is to find a $\Delta$-covered set of ${\cal F}_n$ 
for a non-negative integer, $\Delta$, 
while processing each transaction $t_i$ ($1\leq i \leq n$) only once.
As explained before, PC mining and RC mining approaches for FIM-SD have been established. 
PC mining controls $\Delta$ according to an error parameter, 
$\epsilon$, such that $\Delta\leq \epsilon n$; however, in the worst case, an exponential number of candidate itemsets must be stored 
with this approach. Indeed, it is not feasible for any PC method to solve this problem in $O(k)$ space for a constant $k$.

Hence, we aim to design an RC method by which $\Delta$ is kept as small as possible based on a given value of $k$, which bounds the memory consumption.
Herein, we present a novel RC method to find a $\Delta$-covered set of ${\cal F}_n$ for a given $k$ in $O(k L n)$ time and $O(k)$ space, where $L$ is the maximum length of transactions and $n$ is the end timestamp.

\section{Related work}

Several methods have been proposed for mining condensed representations~\cite{xin05,cheng06,song07,cheng08,boley09,liu12}. 
In this section, we briefly review the state-of-the-art techniques. 

\subsection{Related work on offline mining methods}
Boley {\it et al.} \cite{boley09} proposed an offline algorithm to enumerate {\it $\Delta$-closed sets} from a transaction database.  
This method involves the use of a closure operator, in which each fixed point corresponds to a $\Delta$-closed set, to compute the frequent $\Delta$-closed sets, ${\cal F}_{\Delta}$. The computation time is $ O(N^2 n |{\cal F}_n|)$. 
Unlike $\Delta$-covered sets, ${\cal F}_{\Delta}$ does not satisfy the $\Delta$-deficiency requirement while ${\cal F}_{\Delta}$ is included in any $\Delta$-covered set; this limitation restricts the potential applications of this method. 
The appendix details the relationship between ${\cal F}_{\Delta}$ and any $\Delta$-covered set.

Xin~{\it et al.}~\cite{xin05} introduced another relaxed metric, called the {\it $\delta$-cluster}, which is a $\Delta$-deficient condensed representation based on a relative error metric.
In that study, NP-hardness was applied to find the minimum 
$\delta$-cluster covering the FIs. In addition, a two-pass algorithm, called {\it RP-local}, was introduced to construct a $\delta$-cluster based on a greedy strategy. 
Liu~{\it et al.}~\cite{liu12} improved the efficiency for computing 
the $\delta$-cluster using a {\it CFP-tree}.

Cheng {\it et al.}~\cite{cheng06} introduced a condensed representation, called a {\it $\delta$-tolerance closed set}, and presented a two-pass algorithm to enumerate $\delta$-tolerance closed sets using an FP tree. Unlike $\delta $-clusters, the maximum error is not controlled by the value $\delta$.

\subsection{Related work on online mining methods}

\begin{table*}[t]
	\centering
	\begin{tabular}{c|c|c|c|c}
	 Method & Approach  & Solution &  Accuracy  & Memory \\ \hline
	 CLAIM~\cite{song07} & PC &  Relaxed CFIs & High & Medium\\
	 StreamMining~\cite{jin05} & PC & FIs                         & High      & Exhaustive        \\
     CloStream~\cite{yen11} & NC & Closed FIs                & Exact     & High         \\
     Skip LC-SS~\cite{yamamoto14} & RC & FIs                & Low       & Constant   \\
     IncMine~\cite{cheng08} & PC & SFCIs                        & High   & Medium      \\
	 \hline\hline
     PARASOL (this study) & Hybrid & $\Delta$-covered set & High      & Constant    
	\end{tabular}
		\caption{Summary of online itemset mining methods}
				\label{related_work}
\end{table*}
Recently, several online algorithms for dealing with lossy condensed representations were proposed; several representative algorithms are
summarized in Table~\ref{related_work}.

Song~{\it et al.}~\cite{song07} presented the notion of {\it relaxed closed itemsets} which is a $\Delta$-deficient condensed representation. 
Given an error parameter, $\epsilon$, the frequency range of the itemsets is divided into $\lceil \frac{1}{\epsilon} \rceil$ intervals. 
Then, closed itemsets representing upper or lower bounds for each interval with respect to the inclusion relation were identified. 
By definition, these closed itemsets composed a $\Delta$-covered set. 
Thus, the notion of $\Delta$-covered sets is regarded as a generalization of the relaxed closed itemsets. 
This study also included the development of an online PC approximation algorithm called {\it CLAIM}, which incrementally updates relaxed closed itemsets by computing the {\it drifted} itemsets for each timestamp. However, CLAIM invokes a relatively complex updating process which makes it slow. 

Cheng {\it et al.} \cite{cheng08} proposed an incremental method called {\it IncMine} to compute {\it semi-frequent closed itemsets} (SFCIs). 
IncMine incrementally maintains SCFIs for each segment (i.e., each set of transactions) in two steps: computing the SFCIs in the current segment followed by updating whole set of SFCIs based on the newly computed ones.  
Compared with CLAIM, IncMine utilizes various pruning techniques and efficient data structure. IncMine was implemented in a {\it massive online analysis} (MOA) platform which enables it widely distributed
~\cite{quadrana15}. However, the updating process used by IncMine still requires subsets to be enumerated for each SCFI, resulting in a large time delay in some cases.  
Even worse, it is unknown if the obtained SFCIs are $\Delta$-deficient.

{\it StreamMining}~\cite{jin05} is a PC approximation method that seeks FIs.  Based on an error parameter, $\epsilon$, it deletes  itemsets that are not promising as they have frequencies less than or equal to $\epsilon\times n$. The accuracy of the solution (i.e., the maximum error in the restored frequencies) can be directly controlled by $\epsilon$.
However, this process involves a combinatorial explosion of the FIs causing huge memory consumption. 
{\it CloStream}~\cite{yen11} is a non-constrained (NC) online method that computes closed FIs exactly. 
Its solution is oriented to condensed representation of FIs, while memory efficiency is bounded by the limitation of lossless compression based on closedness. 

To our knowledge, IncMine~\cite{cheng08,quadrana15} is the state-of-the-art PC method for condensed representation mining. 
Unlike IncMine, PARASOL takes RC approximation just in case for processing concept drifts. Then, it can tolerate any busty transaction without memory overflow.  
In usual, it takes the PC approximation to keep the accuracy of its solution high, provided that the solution ensures the $\Delta$-deficiency property. 

Skip LC-SS~\cite{yamamoto14} is an RC method that seeks FIs based on a {\it space-saving} technique~\cite{metwally05}. 
Given a size constant $k$, it can process any transaction with length $L$ in $O(k)$ space and $O(kL)$ time, keeping the top-$k$ itemsets with respect to their frequencies. 
However, Skip LC-SS must increment $\Delta$ by one, whenever a transaction $t_i$ that satisfies $2^{|t_i|} > k$ is processed; 
hence, it suffers from a rapid increase in $\Delta$ in accordance with $k$. 

To mitigate this issue, PARASOL introduces a lossy condensed representation of FIs and takes RC approximation only for processing concept drifts. Thus, it can tolerate bursty transactions without causing memory overflow. In most cases, however, PARASOL uses the PC approximation to keep the accuracy of the solution high provided that the solution exhibits $\Delta$-deficiency.

Compared with the previously developed techniques, the present approach is novel in that is combines PC and RC methods to mine a 
$\Delta$-deficient solution. Although several online algorithms for FIM-SD have been proposed so far, none offer both high accuracy and space efficiency. Unlike the alternative methods, PARASOL avoids sudden and intense memory consumption associated with concept drift.


%

\section{Baseline algorithm for online $\Delta$-covered set mining}

PARASOL mines a $\Delta$-deficient condensed solution based on one-pass approximation settings. 
PARASOL is built on two key techniques: {\it incremental intersection} and {\it minimum entry deletion}, which are described in detail in this section. 

{\it Incremental intersection}~\cite{borgelt11, yen11} is used for computing the closed itemsets. 
It is based on the following cumulative and incremental features of the closed itemsets.
Let ${\cal C}_i$ be the family of closed itemsets in ${\cal S}_i$. 
\begin{theorem}\label{update_them}\rm\cite{borgelt11, yen11}
Given ${\cal C}_i$ at timestamp $i$ and transaction $t_{i+1}$ at timestamp $i+1$, ${\cal C}_{i+1}$ is defined as follows:
\begin{eqnarray}
  {\cal C}_{i} &=& \emptyset ~~ \mbox{for}~~ i = 0 \nonumber \\
  {\cal C}_{i} &=& {\cal C}_{i-1} \cup \{t_{i}\} \cup  \nonumber \\ 
                & & \{\ \beta  ~|~ \beta = \alpha\ \cap\ t_{i}, \beta\neq\emptyset, \alpha\in {\cal C}_{i-1}\  \} ~~ \mbox{for} ~~ i \geq 1  \nonumber 
\end{eqnarray}
\end{theorem}
Theorem \ref{update_them} ensures that ${\cal C}_{i+1}$ can be computed from the intersection of each itemset in ${\cal C}_i$ 
with $t_{i+1}$. 
%
\begin{figure}[t]
	\begin{center}
	  	\includegraphics[width=8.5cm]{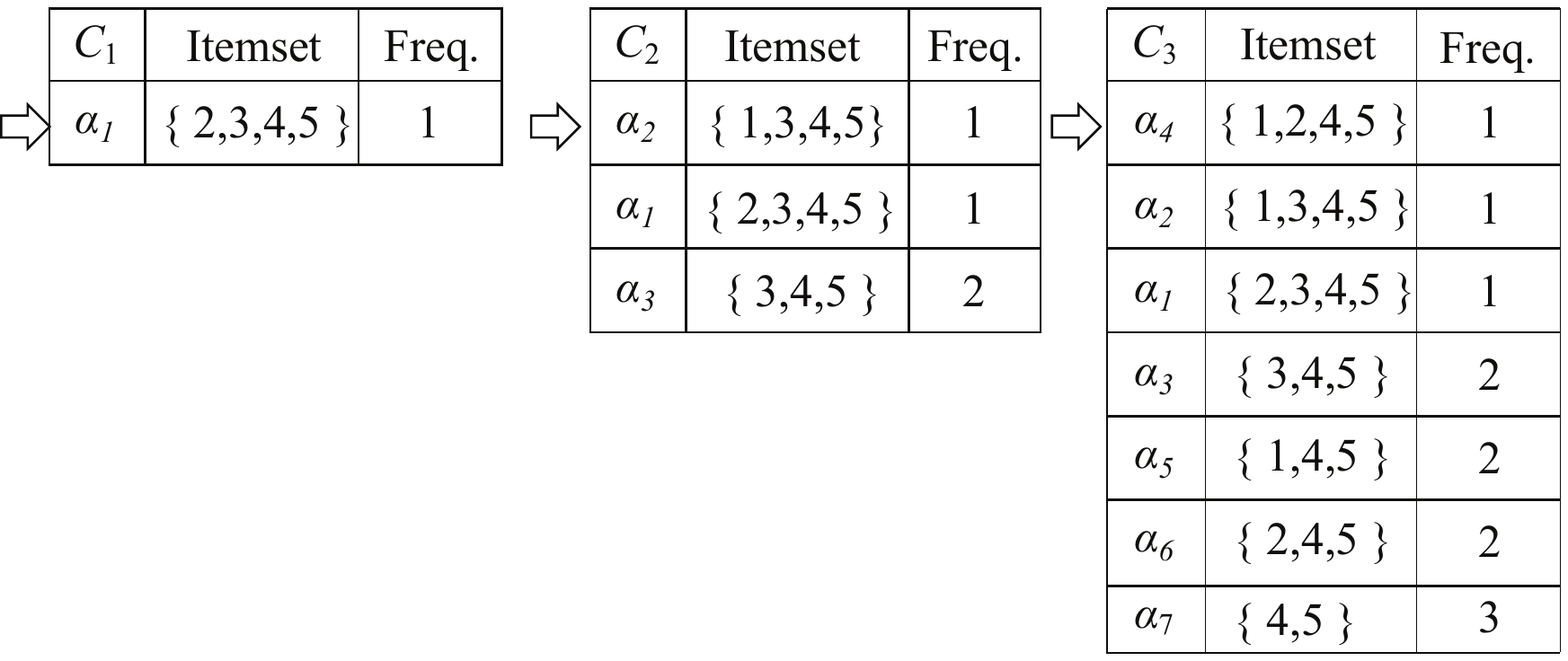}
				  \vspace{-10mm}	
			\caption{Example of incremental intersection}
		\label{inc_inter_ex}
	\end{center}
\end{figure}
\begin{example}\label{ex5}\rm
Let $I = \{1, 2, 3, 4, 5\}$ and the stream ${\cal S}_4^2 = \langle t_1, t_2, t_3, t_4 \rangle$ where
$t_i = I - \{i\}$ for each $i$ $(1\leq i \leq 4)$.
Figure~\ref{inc_inter_ex} illustrates how ${\cal C}_i$ is incrementally generated until $i = 3$. 
${\cal C}_1$ consists of the first transaction.
${\cal C}_2$ newly contains $\alpha_2$ and $\alpha_3$, which correspond to the second transaction 
and its intersection with $\alpha_1$, respectively. 
Finally, ${\cal C}_3$ and ${\cal C}_4$ consist of the seven and fifteen closed itemsets, respectively.
\end{example}

An incremental intersection never unfolds a transaction to intermediate subsets. 
In this sense, it can simply but efficiently compute only the closed itemsets.
On the other hand, lossless compression based on closedness cannot necessarily control an exponential increase in the number of closed itemsets; in the worst case, $\Omega(2^L)$ closed itemsets can be generated. In Example~\ref{ex5}, 
there is a total of fifteen closed itemsets from ${\cal S}_4^2$, which is almost the same as the $2^{L}$ closed itemsets that are attained for $L=4$.

Next, we introduce a space-saving technique ~\cite{metwally05,yamamoto14} for RC approximation, called {\it minimum entry deletion}.
Given a size constant, $k$, RC methods generally sweep the currently stored itemsets and keep only the top-$k$ itemsets;
in other words, the itemset with the lowest frequency is deleted iteratively as long as the number of stored itemsets is greater than 
$k$.
The maximum frequency of the deleted itemsets is maintained as the maximum error, $\Delta$.
 
\begin{figure}[t]
	\begin{center}
				 \vspace{-30mm}	
	  	\includegraphics[width=7.5cm]{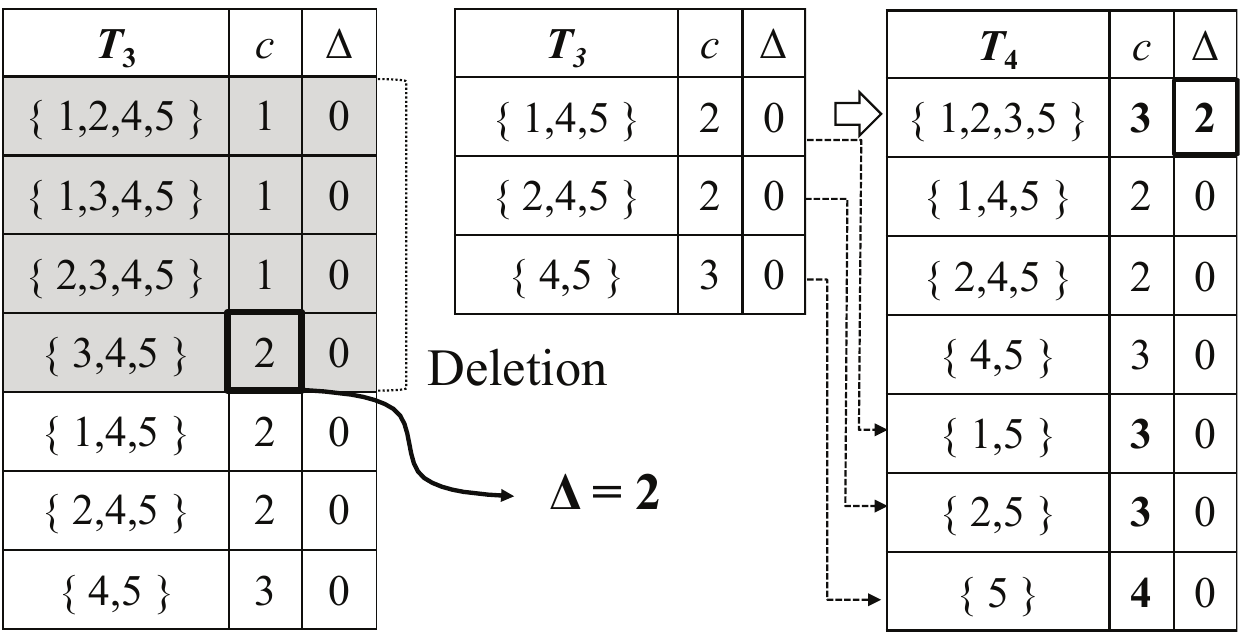}
				  \vspace{-30mm}	
			\caption{Example of minimal entry deletion}
		\label{minEntryFigEx}
	\end{center}
\end{figure}

PARASOL manages two types of information for each stored itemset: its {\it estimated frequency} and {\it maximum error}. 
They are represented as a tuple entry of the form $\langle \alpha, c, \Delta\rangle$, which represents the stored itemset, the frequency count, and the error count, respectively.
The itemset, frequency count, and error count of an entry, $e$, are often denoted by $\alpha_e$, $c_e$, and $\Delta_e$, respectively.
\begin{example}\rm\label{min_entry_fig}
Consider ${\cal S}_4^2$ again. Let $k=3$. 
The number of ${\cal C}_3$ is beyond $k=3$ at timestamp $i=3$.
As shown in Figure~\ref{minEntryFigEx}, we remove four closed itemsets in order of increasing frequency; the highest frequency of those that are removed is two.
The next transaction, $t_4$, is stored in the entry $\langle t_4, 3, 2\rangle$, meaning that the true frequency of $t_4$ is between one and three (i.e., $1\leq sup(t_4, 4)\leq 3$). 
\end{example}
In the following, $T_i$ denotes the collection of entries used at timestamp $i$, 
$k(i)$ denotes the number of entries in $T_i$, and $\Delta(i)$ denotes the maximum error at timestamp $i$.
A {\it minimum entry} in $T_i$ is an entry whose frequency count is the lowest of those in $T_i$. 
The baseline algorithm is described in Algorithm~\ref{com_stream}.
\begin{algorithm}
\caption{The baseline algorithm} \label{com_stream}
\begin{algorithmic}[1]
\Require a size constant ($k$), minimum support threshold ($\sigma$) and data stream (${\mathcal S}_n = \langle t_1,\ t_2,\ \ldots,\ t_n \rangle$).
\Ensure a $\Delta(n)$-covered set of the FIs with respect to $\sigma$.

\State set $i$ as 1 ($i := 1$)  \Comment {$i$ is the current time}
\State $\Delta(i) := 0$ \Comment {$\Delta(i)$ is the maximum error at time $i$}
\State initialize $T_0$
\While {$i \leq n$} read $t_i$
	\State {$T_i := intersect(T_{i-1}, t_i)$}
	\State {$\Delta(i) := delete(T_i)$}
	\State {$i := i + 1$}
\EndWhile
\For {each entry $e$ in $T_n$ such that $c_e > \sigma n$} 
\State {output $e$} \Comment {composing a $\Delta(n)$-covered set}
\EndFor
\end{algorithmic}
\end{algorithm}
\begin{figure}[t]
\begin{algorithmic}[1]
\algrule
\Function{intersect}{$T_i, t_{i+1}$}
\State {initialize $C$} \Comment {$C$: the collection of candidate entries}
	\State {$e := get(t_{i+1}, T_i)$} \Comment{getting the entry $e$ for $t_{i+1}$ in $T_i$}
\If {$e$ is null} 
		\State add the entry $\langle t_{i+1},\ \Delta(i),\ \Delta(i) \rangle$ to $T_i$
	\EndIf
\For {each entry $u$ in $T_i$ such that $\alpha_u \cap t_{i+1} \neq \emptyset$}
	\State {$v := get(\alpha_u \cap t_{i+1}, C)$} 
		\If {$v$ is null} 
		\State {add $\langle \alpha_u \cap t_{i+1},\ c_u + 1,\ \Delta_u \rangle$ to $C$}
	\ElsIf {$c_v < c_u + 1$ } 
		\State{$c_v := c_u + 1$, $\Delta_v := \Delta_u$} 
	\EndIf	  
\EndFor
\For {each entry $r$ in $C$}
	\State {$e := get(\alpha_r, T_i)$}\Comment{getting the entry $e$ for $\alpha_r$}
	\If {$e$ is null} \Comment {no entry for $\alpha_r$}
		\State{add the entry $r$ to $T_i$}
	\Else
		\State{replace the entry $e$ with the entry $r$ in $T_i$}
	\EndIf
\EndFor
\State{return $T_i$} \Comment{corresponding to the next entry table}
\EndFunction\algrule
\end{algorithmic}
\begin{algorithmic}[1]
\Function{delete}{$T_i$}
\While {$k < k(i)$} 
	\State {$m := getMin(T_i)$} \Comment {$m$ is a minimum entry in $T_i$}
	\State {$\Delta(i) := c_m$} \Comment {$c_m$ is the minimum frequency in $T_i$}
	\State {$deleteMin(T_i)$} \Comment {delete a minimum entry in $T_i$}
\EndWhile
\State {return $\Delta(i)$}
\EndFunction
\algrule
\end{algorithmic}
\end{figure}

\begin{figure}[t]
 	\vspace{-25mm}
	\begin{center}
	  	\includegraphics[width=7cm]{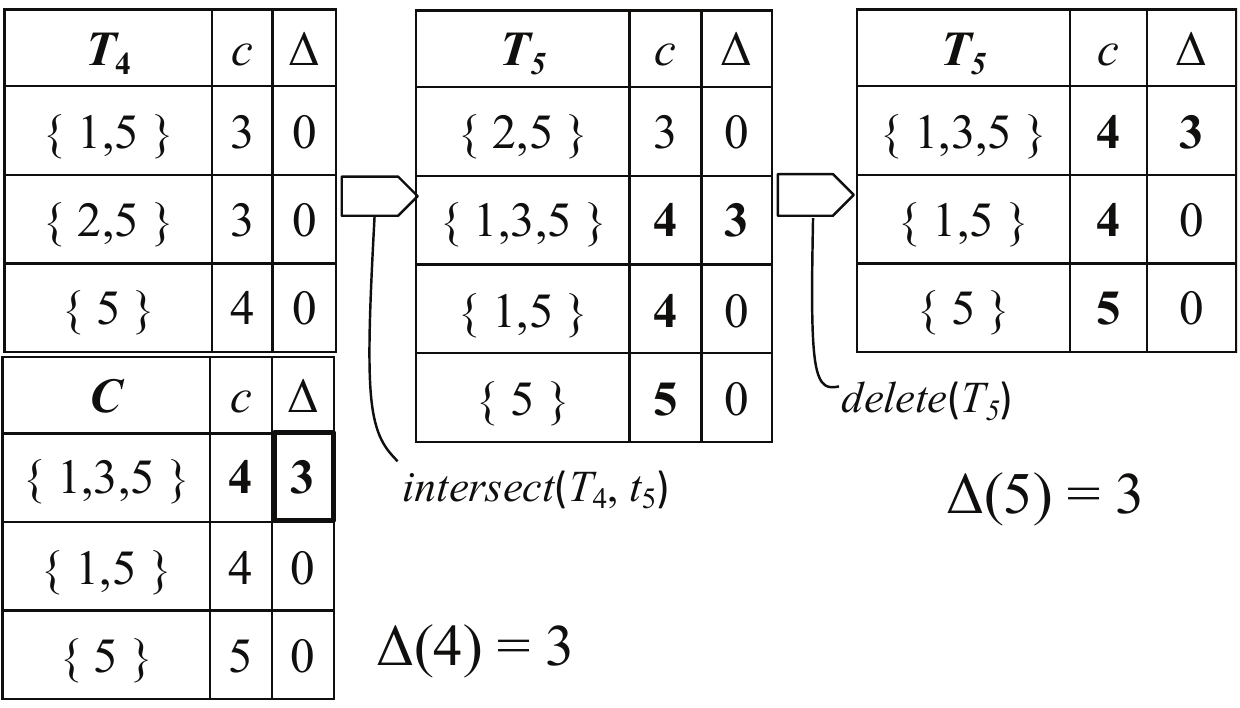}
 	\vspace{-30mm}
			\caption{Example of the baseline algorithm}
		\label{clo_stream_ex}
	\end{center}
\end{figure}

 
The updating process is composed of entry addition 
and entry deletion for each timestamp. 
The addition operation is carried out by the function $intersect(T_{i-1}, t_i)$, which performs the incremental intersection.
The deletion operation is realized by the function $delete(T_i)$, which performs the minimum entry deletion.
%
%
\begin{example}\label{com_stream_ex}\rm
Consider $T_4$ again as shown in Figure~\ref{minEntryFigEx}. 
$T_4$ corresponds to the output of Line~5 in Algorithm~\ref{com_stream} at time $i=4$ for the input stream ${\cal S}_4^2$ and $k = 3$. By the deletion process of Line~6, $\Delta(4)$ becomes three.
Now, we assume that the top three entries in $T_4$ remain as shown in Figure~\ref{clo_stream_ex}.
Next, let the transaction arriving at time $i=5$ be $t_5 = \{1,3,5\}$. 
Then, by the addition process, we first add the entry $\langle t_5, 3, 3\rangle$ to $T_4$, since there is no entry for $t_5$ in $T_4$. 
After that, the intersection of each itemset stored in $T_4$ with $t_5$ is computed and stored in $C$. 
In total, $C$ consists of three entries. 
Then, $T_5$ is obtained by updating $T_4$ with $C$ as shown in Lines~15-22 using the intersect function. 
Finally, one minimum entry, $\langle \{2,5\}, 3, 0 \rangle$, is deleted from $T_5$ to ensure that $k(5) \leq k$.
\end{example}
Next, we clarify the quality of the output. 
\begin{theorem}\label{quality}\rm
Algorithm~\ref{com_stream} outputs a $\Delta(n)$-covered set of the FIs wrt $\sigma$, provided that $\Delta(n) \leq \sigma n$.
\end{theorem}
This implies that for every FI, $\alpha$, $T_n$ must contain an entry, $e$, in which $\alpha$ is $\Delta(n)$-covered 
by $\alpha_e$. To demonstrate this feature, we introduce the notion of a {\it representative entry}.
\begin{definition}\label{restore_def}\rm
Let $\alpha$ be an itemset. Given $T_i$, $T_i^{\alpha}$ is used to denote the set of entries in $T_i$ that contains such an itemset $\beta$ that $\alpha\subseteq \beta$.
If $T_i^{\alpha}$ is not empty, a {\it representative entry}, $r$, for $\alpha$ is defined as 
an entry that has the maximum frequency count of those in $T_i^{\alpha}$ (i.e., $r = argmax_{e \in T_i^{\alpha}} (c_e)$). 
\end{definition}
Then, for every FI, $\alpha$, and a representative entry, $r$, of $\alpha$, 
we claim that $\alpha$ is $\Delta(n)$-covered by $\alpha_r$.
This claim can be explained as shown in Figure~\ref{clo_stream_ex}:
Since $\Delta(5) = 3$, we set $\sigma=0.6$ to ensure that $\Delta(5)\leq \sigma\times 5$.
There exist five FIs $\{1\},\{3\},\{5\},\{1,5\},\{3,5\}$ wrt $\sigma$ in ${\cal S}_5^2$.
Let $\alpha$ be the FI $\{3,5\}$. $T_5^{\alpha}$ uniquely contains the entry $r=\langle \{1,3,5\}, 4, 3\rangle$.
Hence, $r$ is the representative entry of $\alpha$. 
Note that $\alpha = \{3,5\}$ is $1$-covered by $\alpha_r = \{1, 3, 5\}$, since $sup(\{1,3,5\}, 5) = 3$ and $sup(\{3,5\}, 5) = 4$.

First, we prove the following proposition and two lemmas.
\begin{proposition}\label{monotonicity}\rm
Let $e$ be an entry in $T_i$. Then, $e$ is a representative entry for $\alpha_e$.
\end{proposition}
{\it Proof.} 
Suppose that $e$ is not a representative entry. 
Then, there exists another entry, $e'\in T_i$, such that $\alpha_e\subseteq \alpha_{e'}$ and $c_{e} < c_{e'}$.
This contradicts the anti-monotonicity in $T_i$, which is proved by using the mathematical induction, following the below argument. 
First, since $T_1$ contains only one entry, $T_1$ satisfies the anti-monotonicity property.
Assume that $T_i$ satisfies this property (i.e., for every two entries, $e$ and $e'$,
if $\alpha_e\subseteq \alpha_{e'}$ then $c_e \geq c_{e'}$ holds). 
Then, by the cumulative feature of the incremental intersection, the output of the function $intersect(T_i, t_{i+1})$ 
also exhibits anti-monotonicity. 
Moreover, this property is preserved by the deletion process.
Therefore, $T_{i+1}$ also exhibits the anti-monotonicity. $\Box$
\begin{lemma}\label{accuracy}\rm
Let $\alpha$ be an itemset. Given $T_i$, if there exists a representative entry $r$ for $\alpha$ in $T_i$, 
then it holds that
$$c_r - \Delta_r \leq sup(\alpha, i) \leq c_r.$$ 
\end{lemma}
{\it Proof.} We use the mathematical induction. $T_1$ consists of only one entry, $\langle t_1, 1, 0\rangle$. Since $k > 0$, this entry is not deleted. Accordingly, this lemma is true in the case of $T_1$.
Assume that $T_i$ satisfies the claim of the lemma. 
Let $\alpha$ be an itemset whose representative entry, $r_{i+1}$, exists in ${T_{i+1}}$. 
Now, we consider two cases for the existence of a representative entry $r_i$ for $\alpha$ in $T_i$.\\
{\bf Case 1:} $r_i$ exists. 
Based on the assumption, we have $c_{r_i} - \Delta_{r_i} \leq sup(\alpha, i) \leq c_{r_i}$. 
If $\alpha\not\subseteq t_{i+1}$, then $c_{r_{i+1}} = c_{r_i}$ 
because $T_{i}^{\alpha} \supseteq T_{i+1}^{\alpha}$ and $T_{i+1}^{\alpha} \neq \emptyset$ (i.e., the maximum frequency count in $T_{i}^{\alpha}$ is preserved in $T_{i+1}^{\alpha}$ by the minimum entry deletion); thus, since $sup(\alpha, i+1) = sup(\alpha, i)$,  $c_{r_{i+1}} - \Delta_{r_{i}} \leq sup(\alpha, i+1) \leq c_{r_{i+1}}$. 
Otherwise, since $\alpha\subseteq t_{i+1}$, $c_{r_{i+1}} = c_{r_i} + 1$ holds because $T_{i+1}$ contains the entry for $\alpha_{r_i}\cap t_{i+1}$ which has a frequency count of $c_{r_i} + 1$.
Since $sup(\alpha, i+1) = sup(\alpha, i) + 1$, $c_{r_{i+1}} - \Delta_{r_{i}} \leq sup(\alpha, i+1) \leq c_{r_{i+1}}$. 
Finally, since $\Delta_{r_{i}} \leq \Delta_{r_{i+1}}$, we have $c_{r_{i+1}} - \Delta_{r_{i+1}} \leq sup(\alpha, i+1) \leq c_{r_{i+1}}$.
\\
{\bf Case 2:} $r_{i}$ does not exist. This means that $\alpha$ or its supersets never appear or have been deleted before the time $i$. Hence, $sup(\alpha, i)$ is at most $\Delta(i)$, because $\Delta(i)$ is the upper bound of the frequencies of the deleted itemsets. 
Since $r_{i+1}$ exists but $r_i$ does not, $r_{i+1}$ corresponds to $\langle t_{i+1}, 1+\Delta(i), \Delta(i)\rangle$. 
Hence, $\alpha\subseteq t_{i+1}$ holds. 
Accordingly, $1\leq sup(\alpha, i+1)\ \leq 1 + \Delta(i)$ holds. 
Since $c_{r_{i+1}} = 1 + \Delta(i)$ and $\Delta_{r_{i+1}} = \Delta(i)$,
it holds that $c_{r_{i+1}} - \Delta_{r_{i+1}} \leq sup(\alpha, i+1) \leq c_{r_{i+1}}$. \\
In both cases, $T_{i+1}$ also satisfies the claim of the lemma. $\Box$ 
\begin{figure}[t]
 	\vspace{-30mm}
	\begin{center}
	  	\includegraphics[width=7.5cm]{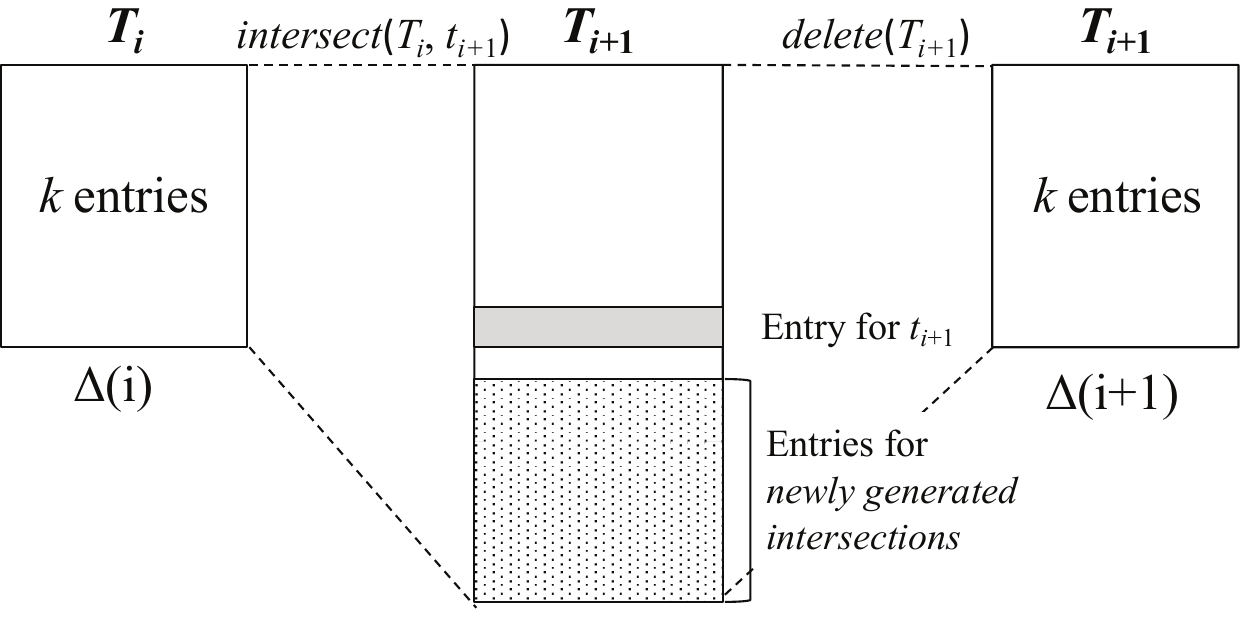}
 		\vspace{-30mm}
			\caption{Sketch of updating $T_i$ to $T_{i+1}$ with $t_{i+1}$}
		\label{intuition_theorem2}
	\end{center}
\end{figure}
\begin{lemma}\label{exist_representative}\rm
For every itemset, $\alpha$, such that $sup(\alpha, i) > \Delta(i)$, there exists a representative entry $r\in T_i$ for $\alpha$. 
\end{lemma}
{\it Proof.} We also use the mathematical induction. 
The lemma is obviously true in the case of $T_1$ 
since $\Delta(1) = 0$ holds. 
Assume that $T_i$ satisfies the claim of this lemma;
we show the induction step by contradiction. 
Assume that $\exists \alpha$ such that $sup(\alpha, i+1) > \Delta(i+1)$ but there is no representative entry for $\alpha$ in $T_{i+1}$. 
We can consider two cases: whether or not a representative entry, 
$r$, has existed in $T_{i+1}$ before the deletion operation (i.e., 
in the intermediate $T_{i+1}$ at the moment when the incremental intersection has been completed, as depicted in the middle $T_{i+1}$ in Figure~\ref{intuition_theorem2}).\\
{\bf Case 1:} such an $r$ exists in $T_{i+1}$ but was deleted.  
By Lemma~\ref{accuracy}, we have $c_r - \Delta_r \leq sup(\alpha, i+1) \leq c_r$.
Since $r$ was deleted, $c_r\leq \Delta(i+1)$ should hold. 
Thus, $sup(\alpha, i+1)\leq \Delta(i+1)$ also holds. However, this is a contradiction.\\
{\bf Case 2:} such an $r$ does not exist in $T_{i+1}$. Here, $T_{i+1}$ captures the entry table before the deletion operation. 
Thus, $T_{i}^{\alpha} \subseteq T_{i+1}^{\alpha}$ holds. 
Since there is no representative entry for $\alpha$ in $T_{i+1}$, it holds that $T_{i+1}^{\alpha} = \emptyset$ and $T_{i}^{\alpha} = \emptyset$. 
Note that $T_{i+1}$ should contain the entry for $t_{i+1}$.
Hence, $\alpha\not\subseteq t_{i+1}$ because $T_{i+1}^{\alpha} = \emptyset$.
Therefore, $sup(\alpha, i) = sup(\alpha, i+1)$.
Since $sup(\alpha, i+1) > \Delta(i+1)$ and $\Delta(i+1) \geq \Delta(i)$, 
it should hold that $sup(\alpha, i) > \Delta(i)$. 
Then, by the assumption for $T_i$, there should exist 
a representative entry for $\alpha$ in $T_i$. 
However, this contradicts the conclusion that $T_i^{\alpha} = \emptyset$.
\\
In both cases, there is a contradiction. Therefore, for every $\alpha$ such that $sup(\alpha, i+1) > \Delta(i+1)$,
there exists a representative entry for $\alpha$ in $T_{i+1}$. $\Box$ 

Now, we prove Theorem~\ref{quality} as follows: \\
{\it Proof of Theorem~\ref{quality}.}
Suppose that the condition of $\sigma$ (i.e., $\Delta(n)\leq \sigma n$) is satisfied.
Then, by Lemma~\ref{exist_representative}, for every FI $\alpha$ wrt $\sigma$,
there exists a representative entry, $r$, in $T_n$ for $\alpha$.
By Lemma~\ref{accuracy}, $c_r-\Delta_r \leq sup(\alpha, n) \leq c_r$ holds.
By Proposition~\ref{monotonicity}, $r$ is also a representative entry for $\alpha_r$.
Thus, we have $c_r-\Delta_r \leq sup(\alpha_r, n) \leq c_r$.
Hence, $sup(\alpha, n)\leq sup(\alpha_r, n) + \Delta_r$ holds. 
Since $\alpha_r\supseteq \alpha$, $\alpha$ is $\Delta_r$-covered by $\alpha_r$.
Since $\Delta_r\leq \Delta(n)$, we can claim that for every FI $\alpha$, there exists an entry, $r$ ,
such that $\alpha$ is $\Delta(n)$-covered by $\alpha_r$. 
Note that $c_r > \sigma n$ holds because $sup(\alpha, n)>\sigma n$.
Hence, $r$ is contained in the output because of Lines 9-11 in Algorithm~\ref{com_stream}. $\Box$ 

%

%

Next, we clarify the complexity of the baseline algorithm.
Incremental intersection generates at most $k(i) + 1$ new entries for the intersections of the entries 
in $T_i$ with $t_{i+1}$ as well as $t_{i+1}$ itself; these new entries are added to $C$ in Lines~5 and 10 of the $intersect$ function. 
Thus, the total number of stored entries is at most $ 2 \times k(i) + 1$, and $k(i)$ is always bounded 
by the size constant $k$.
Hence, the complexity of the baseline algorithm can be described as follows:
\begin{theorem}\label{complexity}
For a size constant $k$ and the maximum transaction length $L$, Algorithm~\ref{com_stream} processes every transaction in 
$O(kL+k\log k)$ time and $O(k)$ space. 
\end{theorem}
{\it Proof.}
First, the number of stored entries is at most $2k + 1$ for every time point. 
Thus, the space required to store them is $O(k)$.
In addition, it takes at most $O(2L)$ time to compute the intersection of a stored itemset with a transaction. 
Thus, the incremental intersection is performed in $O(kL)$ time.
It takes at most $O(\log k)$ time to delete a minimum entry and add a new entry by using a heap structure. 
Hence, the total time required to update the table is at most $O(kL + k\log k)$. $\Box$

Based on Theorems~\ref{quality} and~\ref{complexity}, Algorithm~\ref{com_stream} can extract a $\Delta$-covered set of the FIs wrt $\sigma \geq \frac{\Delta}{n}$ for an integer $\Delta$ and can process each transaction in $O(kL)$ time and $O(k)$ space for a size constant $k$ ($k < 2^L$).
Therefore, Algorithm~\ref{com_stream} satisfies the problem setting presented in Section~2.1.

\section{Improvement of the baseline algorithm}
In this section, we introduce two key improvements to the baseline algorithm.

\subsection{Unifying PC and RC approximations}

We observe in Figure~\ref{moti_ex} that a sudden and significant memory consumption may occur as a temporary phenomenon.  
In other words, the PC approximation is feasible unless all of the memory is consumed during this short period. 
Hence, it is reasonable to embed a size constant, $k$, into the PC approximation and normally perform 
the deletion operation using an error parameter, $\epsilon$, but switch to an RC approximation only when 
the number of used entries exceeds $k$. 
We call this unified approximation scheme {\it PARASOL} (an acronym for Parameter- and Resource-constrained Approximation for Soft and Lazy mining).   

Unlike the existing PC approximation methods, PARASOL can process any bursty transactions in $O(kL)$ time without running out-of-memory. 
In contrast, PARASOL can process ordinary transactions without filling all $k$ possible entries while ensuring that the error ratio, 
$\frac{\Delta(n)}{n}$, is less than or equal to the error parameter, 
$\epsilon$. 
This is unlike to the baseline algorithm based on the RC approximation.
Given $k$ and $\epsilon$, PARASOL can be realized just by replacing the $delete$ function in 
Algorithm~\ref{com_stream} with the following function:
\begin{algorithmic}[1]
\algrule
\Function{parasol\_delete}{$T_i$}
	\State {$m := getMin(T_i)$} \Comment {$m$ is a minimum entry in $T_i$}
	\While {$k < k(i)$ or $c_m \leq \epsilon \times i$} 
		\State {$\Delta(i) := c_m$} \Comment {$c_m$ is the minimum frequency in $T_i$}
		\State {$deleteMin(T_i)$} \Comment {delete a minimum entry in $T_i$}
		\State {$m := getMin(T_i)$}
	\EndWhile
\State {return $\Delta(i)$}
\EndFunction
\algrule
\end{algorithmic}

If there is no deletion for the RC approximation (i.e., $k(i) \leq k$ for every time $i$), PARASOL ensures that $\Delta(i) \leq \epsilon \times i$; otherwise, $\Delta(i)$ can become greater than $\epsilon \times i$. Even in this case, the error ratio, $\frac{\Delta(n)}{n}$ gradually converges to $\epsilon$ as the RC approximation is only required for a short period. 
This {\it self-sustained recovery} of the error ratio is a unique and advantageous characteristic of PARASOL and will be 
empirically demonstrated later. 

\subsection{$\Delta$-compression}
Note that the solution to this problem setting is not unique: there can exist many $\Delta$-covered sets of 
${\cal F}$.
However, it is useful to extract a concise $\Delta$-covered set.
Here, we propose a fast post-processing technique to reduce the original output to a more concise one based on the maximum error, $\Delta(n)$.

We explain the reasoning using $T_5$ (the right-most table in Figure~\ref{clo_stream_ex}) in Example~\ref{com_stream_ex}.
Since $\Delta(5) = 3$, the three itemsets stored in $T_5$ compose a $3$-covered set of the FIs. 
Note that if $\{5\}$ was deleted, the two remaining itemsets still compose a $3$-covered set, 
since $\{5\}$ is $1$-covered by $\{1, 5\}$.
Hence, we can delete an entry for any such itemset that is $\Delta(n)$-covered by another in $T_n$.
In exchange for deleting $\{5\}$, we update the entry for $\{1, 5\}$ by $\langle \{1, 5\}, 5, 1\rangle$ (i.e., incrementing
the frequency and error counts by one). 
In this way, the frequency of $\{5\}$ can be restored from the entry for $\{1, 5\}$.  

Note here that the frequency count of entry $e$ is 
not the exact frequency but rather an estimated one (i.e., $c_e$ is not necessarily equal to $sup(\alpha_e, n)$). 
Therefore, it is difficult to identify all pairs of itemsets that satisfy the $\Delta(n)$-covering relationship in $T(n)$. 
However, some of them can be detected based on the following proposition:
\begin{proposition}\rm\label{compression_prop}
Let $e_{1}$ and $e_{2}$ be two entries in $T(n)$. 
Then, $\alpha_{e_1}$ is $\Delta(n)$-covered by $\alpha_{e_2}$ if it holds that $\alpha_{e_1}\subseteq\alpha_{e_2}$ and $c_{e_{1}} \leq c_{e_{2}} - \Delta_{e_{2}} + \Delta(n)$.
\end{proposition}
{\it Proof.} 
By Lemma~\ref{accuracy}, we have $sup(\alpha_{e_1}, n)\leq c_{e_1}$ and $c_{e_2}-\Delta_{e_2}\leq sup(\alpha_{e_2}, n)$. 
Since $c_{e_1}\leq c_{e_2}-\Delta_{e_2} + \Delta(n)$, it holds that $sup(\alpha_{e_1} n)\leq sup(\alpha_{e_2}, n) + \Delta(n)$. $\Box$

Based on Proposition~\ref{compression_prop}, we proposed a post-processing technique, called {\it $\Delta$-compression}, that incrementally deletes the entry for a $\Delta(n)$-covered itemset from $T(n)$, as described in Algorithm~\ref{d_comp}. 
\begin{algorithm}
\caption{$\Delta$-compression} \label{d_comp}
\begin{algorithmic}[1]
\Function{compress}{$T_n$}
\While{$\exists e_1, e_2\in T_n$ s.t. $\alpha_{e_1}\subseteq\alpha_{e_2}$ and \\
		\hspace{14mm} $c_{e_1} \leq c_{e_2} - \Delta_{e_{2}} + \Delta(n)$}
	    \State {$\Delta_{e_2} := \Delta_{e_2} + c_{e_1} - c_{e_2}$}
		\State {$c_{e_2} := c_{e_1}$}
		\State {delete $e_1$ from $T_n$}
	\EndWhile	
\EndFunction
\end{algorithmic}
\end{algorithm}
In exchange for deleting $e_1$, $\Delta_{e_2}$ is updated by increasing it by the difference between $c_{e_1}$ and $c_{e_2}$, 
and $c_{e_2}$ is updated to $c_{e_1}$ in Lines~4 and~5.

A simple implementation of Algorithm~\ref{d_comp} is that for each entry $e_1$, we search $T(n)$ for a corresponding $e_2$.
This can be achieved in $O(k^2)$ time, since $|k(n)| \leq k$. 
   
\section{data structure}

In this section, we address the issue of the data structure that is needed to efficiently realize the three key operations of the proposed algorithm: incremental intersection, deletion, and $\Delta$-compression. Of these, incremental intersection incurs the majority of the computational cost. 
This operation imposes the traverse of every entry, $e$, to compute its intersection with a transaction, $t_i$, for each time~$i$.
This computation is often redundant: for example, 
if $\alpha_e$ has no common items with $t_i$, it is redundant to compute the intersection of $\alpha_e$ with $t_i$.

Based on this observation, Yen {\it et al}~\cite{yen11} proposed an indexing data structure, called {\it cid\_list}, 
corresponding to the vertical format of the stored itemsets: 
for each item, $x$, $cid\_list(x)$ maintains the indexes of the entries corresponding to the itemsets that contain $x$.
Using $cid\_list$, we can focus only on the entries whose indexes are contained in $\bigcup_{x\in t_i} cid\_list(x)$ and compute their intersections
with $t_i$.  
However, the computational cost of updating $cid\_list$ is relatively high: 
the entire $cid\_list$ is dynamically changed by addition and deletion operations. 
This overhead becomes especially high for {\it dense} datasets since most itemsets stored in $T_i$ have some of the same items as $t_i$.

Borgelt {\it et~al}~\cite{borgelt11} proposed a fast two-pass FIM method, called {\it ISTA}, based on incremental intersection.
In this implementation, the prefix tree (as well as patricia) was introduced to efficiently maintain $T(i)$ 
and perform the incremental intersection.
Although it is reasonable to represent $T(i)$ with such a concise data structure, it is not directly applicable 
in the one-pass approximation setting that is used here. 
For example, we cannot use the item frequency as static information, while it is available in the transactional database that allows multi-pass scanning. 
This information is crucial to constructing a compact trie by sorting items in a pre-processing step. 
Note that the trie size can be directly affected by the order of (sorted) transactions. Indeed, ISTA constructs the trie with 306 nodes by treating it as a retail problem (No. 3 in Table~\ref{static_table}), compared to 244,938 nodes when the pre-processing technique is not applied.
Besides, in the context of the SD that emerges concept drift, it is not appropriate to assume a static distribution.

Thus, we consider a novel data structure that is more suitable for our proposed algorithm: it is designed to prune
redundant computations in incremental intersections and 
quickly access the minimum entries and $\Delta$-covered
entries as required for PARASOL deletion and $\Delta$-compression.

\subsection{Weeping tree}

In this paper, we propose a variation of the {\it binomial spanning tree}~\cite{johnsson89,chang05}, called {\it weeping tree}, in which a collection of entries, $T(n)$, can be represented in a {\it binary $n$-cube} as follows:
Let $e$ be an entry in $T(n)$. 
By Theorem~\ref{update_them}, $\alpha_e$ corresponds to the intersection of a certain set $S$ of transactions.
This set can be represented as a binary address $(x_{1},x_{2},\ldots, x_{n})$ where each $x_{j}$ is one if $S$ contains $t_j$ and zero otherwise.
Every $\alpha_e$ has its own binary address. 
Thus, $T(n)$ can be represented as a set of binary addresses, denoted by $V(n)$, each of which
identifies $\alpha_e$ for each entry $e\in T(n)$.
Each binary address can be described by an integer, $x = \Sigma_{j=1}^n (x_{j}\times 2^{n-j})$. 
Then, $V(n)$ corresponds to a subset of $\{1, \ldots, 2^n-1\}$.

Let $x$ and $y$ be two integers ($1\leq x, y \leq 2^n-1$) with the addresses 
$(x_{1},x_{2},\ldots, x_{n})$ and $(y_{1},y_{2},\ldots, y_{n})$, respectively.
$p(x)$ is the position in $x$ that satisfies two conditions:  $x_{p(x)} = 1$ and $x_{j} = 0$ for each $j$ $(p(x)+1\leq j\leq n)$.
In other words, $p(x)$ is the least significant set bit of $x$.
We say that {\it $x$ covers $y$} if $y_j = x_j$ for each $j$ where $1\leq j\leq p(x)$.
For example, if $x = 12$ and $y = 15$ in $4$-cube with addresses $(1100)$ and $(1111)$ then, $p(x) = 2$, $y_1 = x_1$, and $y_2 = x_2$ so $x$ covers $y$.
We assume that zero covers every integer.

Now, we define the binomial spanning tree of $V(n)$, following the notion in the literature~\cite{chang05}.
\begin{definition}[Binomial spanning tree]
Let $V(n)$ be a subset of $\{1, \ldots, 2^n-1\}$, $x$ be an integer such that $0\leq x\leq 2^n-1$,
and $C(x)$ be the set of integers in $V(n)$ each of which is covered by $x$. 
The binomial spanning tree of $V(n)$ is the tree in which the root node, $r$, is zero and the other nodes are $V(n)$. 
The children of each node, $x$, correspond to the following set: 
$$\{ y~|~y\in C(x)\ and \not\exists y'\in C(x)\ s.t.\ y \neq y'\ and\ y \in C(y')\}.$$
The siblings $y^{(1)}, y^{(2)}, \ldots, y^{(m)}$ are sorted in the descending order 
(i.e., $y^{(j)}$ is the left sibling of $y^{(j+1)}$). 
We say that a non-root node, $y$, is a {\it descendant} of node $x$ if $y\in C(x)$, a {\it precursor} of $x$ if $y\not\in C(x)$ and $y > x$, and a {\it successor} of $x$ if $y < x$.
The precursors and successors of $x$ are denoted at $P(x)$ and $S(x)$, respectively.
\end{definition}
The {\it weeping tree} at time $n$, denoted by $W(n)$, is the binomial spanning tree of $V(n)$ obtained by associating 
each node, $x$, with its corresponding entry $e$ (i.e., $\alpha_e$ is the intersection of the transactions indicated by 
the address $x$). 

\begin{figure}[h]
	\vspace{-3mm}
	\begin{center}
		\includegraphics[width=8.5cm]{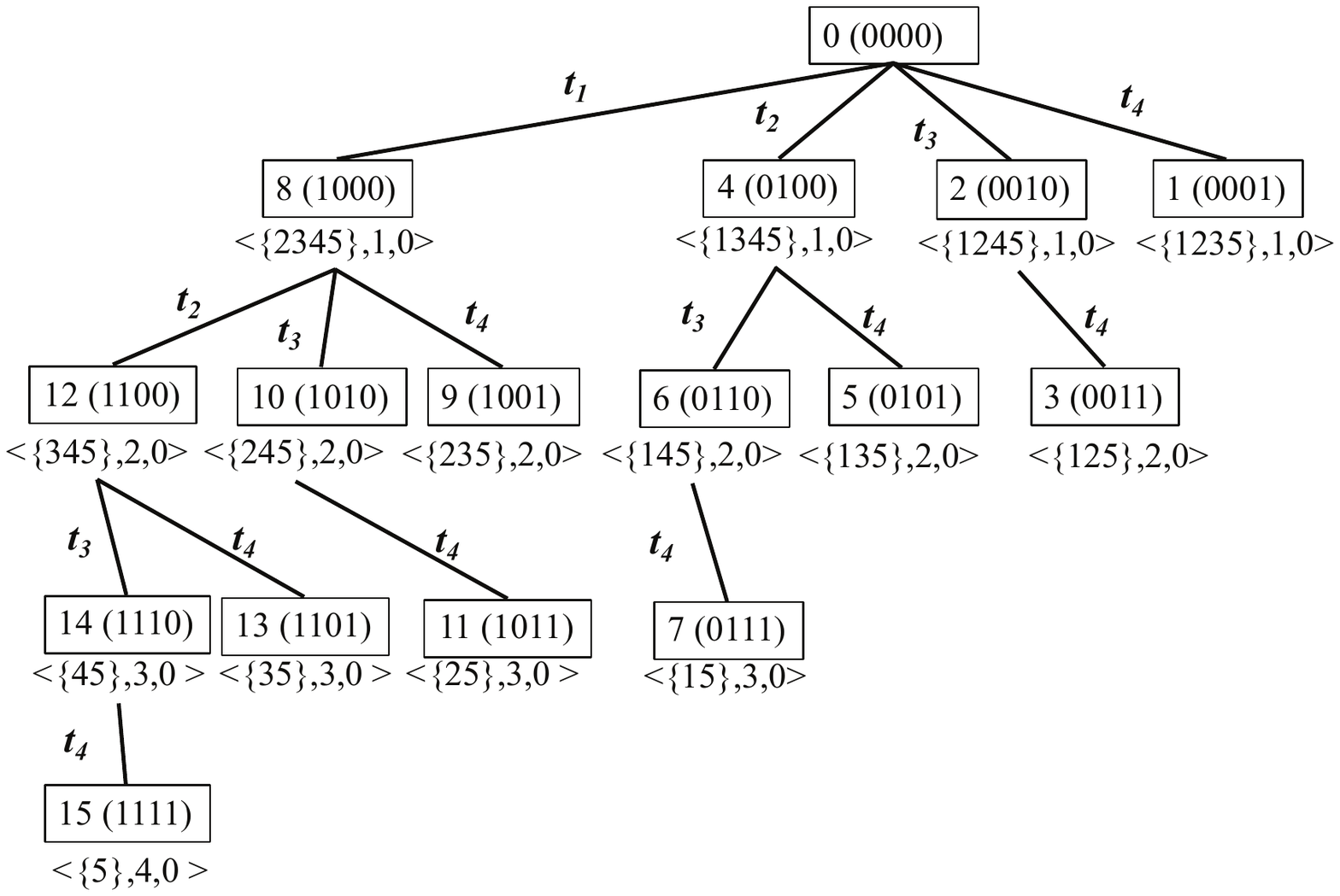}
		\caption{The weeping tree $W(4)$ wrt ${\mathcal S}_4^2$}
		\label{weep_prefix_fig}
	\end{center}
\end{figure}

\begin{example}\label{weep_prefix_ex}
Consider again the stream ${\cal S}_4^2$ in Example~\ref{ex5}.
Let $\epsilon$ and $k$ be 0.2 and 15, respectively.
PARASOL uses the 15 entries in $T(4)$ for all closed itemsets (i.e., $V(4) = \{1, 2,\ldots, 15\}$).
The corresponding weeping tree, $W(4)$, is described in Figure~\ref{weep_prefix_fig}.
Each node, $x$, is associated with its own entry, $e$. 
For example, the node $x = 6$ $(0110)$ corresponds to the entry for $\alpha = t_2\cap t_3$, i.e., $\alpha = \{1, 4, 5\}$. 
Note also that $C(6) = \{7\}$, $P(6) = \{8, 9,\ldots, 15\}$ and $S(6) = \{1, 2,\ldots, 5\}$. 
\end{example}

One crucial feature of the weeping tree is that it captures inclusion relationships among the stored itemsets in $T(n)$. 
\begin{proposition}\label{inclusion_prop1}
Let two nodes $x$ and $y$ associated with two entries $e_x$ and $e_y$. 
If $x$ covers $y$, then $\alpha_{e_y}\subseteq \alpha_{e_x}$
\end{proposition}
{\it Proof.} Since $x$ covers $y$, the address of $y$ can be written as $(x_1, \ldots, x_{p(x)}, y_{p(x)+1}, \ldots, y_n)$.
Accordingly, $\alpha_{e_y}$ is written as $(\bigcap_{x_j = 1, 1\leq j \leq p(x)} t_j)\ \cap\ I$
where $I = \bigcap_{y_j = 1, p(x)+1\leq j \leq n} t_j$.
Since $p(x)$ is the position of the least significant set bit, 
$\alpha_{e_x}=\bigcap_{x_j = 1, 1\leq j \leq p(x)} t_j$ holds, followed by
$\alpha_{e_y} = \alpha_{e_x}\cap I$. $\Box$

Proposition~\ref{inclusion_prop1} has three useful implications. 
First, it is applicable for pruning the intersection computations.
Suppose that during the updating process at time $i$, an entry, $e$, is found such that $\alpha_e \subseteq t_{i}$. 
Since every descendant of $e$ must be included in $t_{i}$, 
it is not necessary to compute the intersections
for these descendants. 
Proposition~\ref{inclusion_prop1} also implies that every minimum entry must be located in the shallowest layer 
in the tree due to the anti-monotonicity of $T(i)$. 
This feature is useful for the minimum entry deletion (in practice, it is reasonable to use min-heap for the shallowest layer).

Finally, Proposition~\ref{inclusion_prop1} is applicable to the pairwise checking involved in $\Delta$-compression.
Suppose that we found a parent entry, $e_p$, and its child entry, $e_c$ such that 
$C_{e_c} \leq C_{e_p}-\Delta_{e_p} + \Delta(n)$. 
Then, $\alpha_{e_c}$ must be $\Delta(n)$-covered by $\alpha_{e_p}$ according to Proposition~\ref{compression_prop}.
Hence, a quick check can be done to determine if each child is $\Delta(n)$-covered by its parent. 
Note that a brute-force approach requires $O(k^2)$ time for $\Delta$-compression while the quick pairwise checking can be completed in $O(k)$ time.
Thus, it is useful as a pre-processing step preceding $\Delta$-compression.

\subsection{Weeping tree updating}

Here, we explain how to incrementally update $W(i)$ at each time $i$.
Suppose that an itemset, $\alpha$, is newly stored in $W(i+1)$.
An address $(x_1, \ldots, x_i, 1)$ is then assigned to $\alpha$,
where $(x_1, \ldots, x_i)$ is the address of the node, $r$,
in $W(i)$ that corresponds to a representative entry of $\alpha$ if such an $r$ exists; if there is no such node, $r$ is the root node. 
Thus, the entry for $\alpha$ is newly located as a child of 
node $r$. 

\begin{example}\label{weeping_tree_update}
Consider Example~\ref{weep_prefix_ex} again.
Now, let $\alpha$ be the itemset $\{2,5\}$ that is newly added in $W(4)$.
There exists a representative entry $r = \langle \{2, 4, 5\}, 2, 0 \rangle $, for $\alpha$ (see Figure~\ref{minEntryFigEx}) 
in $T(3)$.
Since $r$ is given the address $(101)$ at time $i = 3$, the address of $\alpha$ becomes $(1011)$, and the entry for $\alpha$ is located as a child of $r$.
\end{example}

Now, we interpret the meaning of the address assigned to each entry.
Let $e$ be an entry of node $x$ with the address $(x_1,\ldots,x_n)$. We denote the least and greatest significant bit sets of $x$ as $p(x)$ and $q(x)$, respectively. 
Then, $e$ is written as $\langle \alpha, \Delta(q(x)-1) + B, \Delta(q(x)-1)\rangle$,
where $\alpha = \bigcap_{x_{j}=1, q(x)\leq j\leq p(x)} t_j$, $\Delta(q(x)-1)$ is the maximum error at time $q(x)-1$ and 
$B$ is the bit count of $x$ (See Figure~\ref{relation_entry_address}). This observation leads to the following proposition: 
\begin{figure}[t]
		 		\vspace{-50mm}
 	\begin{center}
		\includegraphics[width=9cm]{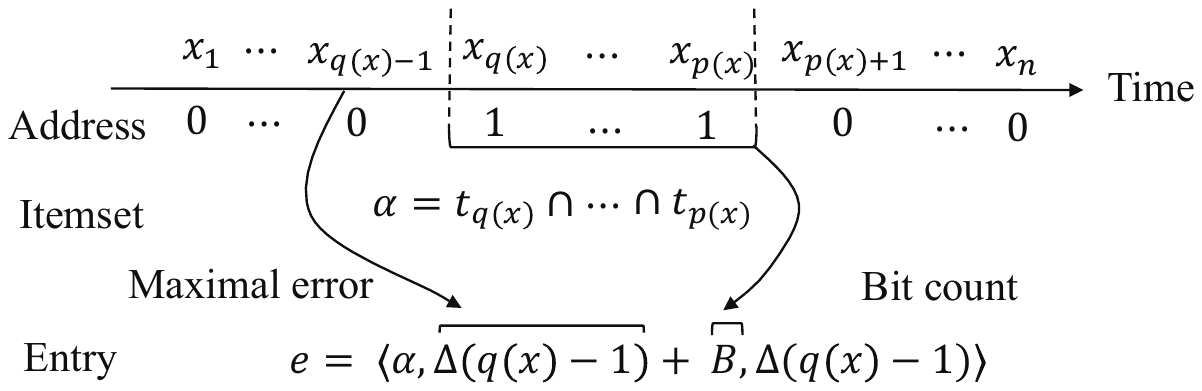}
		 		\vspace{-50mm}
			\caption{Relation between address and entry}
		\label{relation_entry_address}
	\end{center}
\end{figure}
\begin{proposition}\label{inclusion_prop2}
Consider two nodes, $x$ and $y$, of entries $e_x$ and $e_y$, respectively, in a weeping tree.
If $\alpha_{e_y}\subseteq \alpha_{e_x}$, then $y$ is either a descendant or a precursor of $x$.
\end{proposition}
{\it Proof.} 
We derive a contradiction in the case that $\alpha_{e_y}\subseteq \alpha_{e_x}$ and $y < x$. 
Let $v$ be the address obtained by the bitwise OR operation between $x$ and $y$. 
Since $y < x$ and $x\leq v$, $y < v$. 
We write $v$ as $(v_1, \ldots, v_n)$ and denote by $\alpha_v$ the intersection $\bigcap_{v_j = 1, 1\leq j\leq n} t_j$.
Since $\alpha_{e_y}\subseteq\alpha_{e_x}$, we have $\alpha_v=\alpha_{e_y}$. 
Thus, node $v$ should not appear in the tree, since its duplicate never occurs in $T(n)$. 
Without losing generality, this implies that $v$ has been deleted at some time, which is referred to as time $m$.
Accordingly, the tree never contains such a node, $u$, 
with an address $(u_1,\ldots, u_n)$ such that
$q(u) = q(v)$; moreover, the bit count of the $m$-prefix 
$(u_1,\ldots, u_m)$ is lower than the bit count of the 
$m$-prefix $(v_1, \ldots, v_m)$.
This is because every node with such an address has been deleted at time $m$, along with $v$ (i.e., $u$ has a lower 
frequency count than $v$). 
Hence, no node can have an address in which the $m$-prefix matches 
$(u_1, \ldots, u_m)$ or $(v_1, \ldots, v_m)$.
Next, we consider the address of node $x$. 
Since $y < x$ and $v$ is obtained by a bitwise OR operation between $x$ and $y$, we have $q(x) = q(v)$.
In addition, $(x_1, \ldots, x_m)$ is either equal to $(v_1, \ldots, v_m)$ or has a lower bit count 
than $(v_1, \ldots, v_m)$. 
Hence, $x$ should not appear in the tree. 
This is a contradiction. $\Box$     

For example, consider node $6$ for the itemset $\{1,4,5\}$ 
as shown in Figure~\ref{weep_prefix_fig}. 
There are three nodes $7, 14, 15$ that have subsets of this itemset and each of these nodes is either a descendant or a precursor of the node $6$. 

Proposition~\ref{inclusion_prop2} is useful for pruning the computation for incremental intersection.
Suppose that for some entry $C$, $t_{i} \subset \alpha_C$ holds. Thus, the intersection computations for every successor of $C$ with $t_i$ can be skipped as they do not store any subset of $t_i$. 
\begin{figure}[h]
 	\vspace{-25mm}
	\begin{center}
	  	\includegraphics[width=7cm]{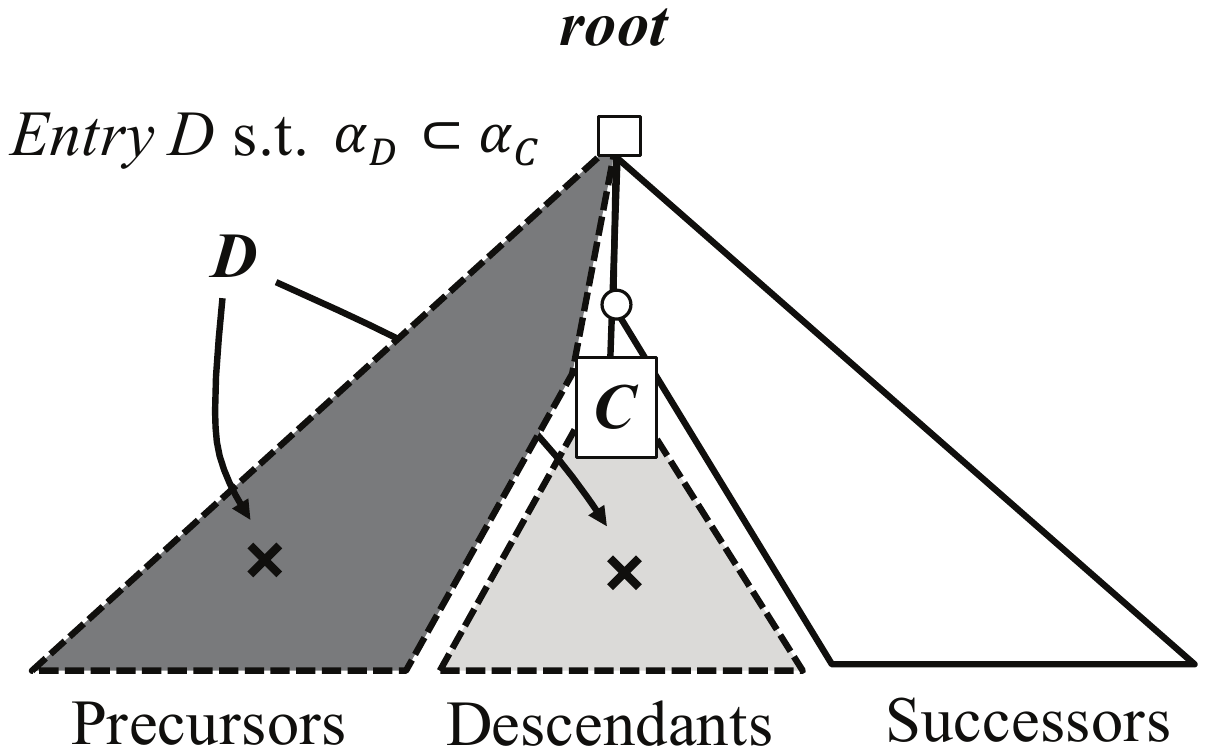}
		 		\vspace{-30mm}
			\caption{Reasoning for Proposition~\ref{inclusion_prop2}}
		\label{weeping_feature}
	\end{center}
\end{figure}

The weeping tree can be used to perform the incremental intersection by traversing the weeping tree in a depth-first, left-to-right manner.  
Algorithm~\ref{weep_tree_const} sketches the process for updating a node $x$ in $W(i)$ with an itemset $E$. 
Note that $E$ is initially a transaction. 

In the algorithm, a node, $x$, is identified with its associated entry, $e_x$. Here, $\alpha_x$, $c_x$ and $\Delta_x$ 
are the itemset, frequency, and error count of $e_x$, respectively.
Given the transaction $t_{i+1}$ and $W(i)$, the next tree, 
$W(i+1)$, is obtained by calling the function 
$update(root, t_{i+1}, W(i))$.
Note that $W(0)$ is defined as the initial tree consisting of the $root$ node.
\begin{algorithm}
\caption{Incremental intersection with weeping tree} \label{weep_tree_const}
\begin{algorithmic}[1]
\Function{update}{$x$, $E$, $W(i)$}
\For{each child $y$ of $x$ in order from left to right}
	\State $I := \alpha_y \cap E$ 
	\If {$|I| = |\alpha_y|$} \Comment{ Case (1) }
		\For{each descendant $z$ of $y$}
			\State {increment $c_z$ by one}
		\EndFor
	\ElsIf {$I \neq \emptyset$} \Comment{ Case (2) }
		\State{$W(i) := update(y, I, W(i))$}	
	\EndIf
	\If {$ |I| = |E|$} \Comment{ Case (3) }
		\State{break the loop}
	\EndIf	
\EndFor 
\If {there is no entry for $E$ in $W(i)$}
 	\State {create $\langle E,\ c_x + 1,\ \Delta_x \rangle$} and 
	\State {add it to the right-most child of $x$}
\EndIf	
\State {return $W(i)$}
\EndFunction
\end{algorithmic}
\end{algorithm}

%
A few characteristics of the $update(x, E, W(i))$ algorithm should be noted:
\begin{itemize}
\item Line~4 means that $\alpha_y\subseteq E$.
By Proposition~\ref{inclusion_prop1}, the descendants of $y$ are included by $E$. 
Thus, the frequency count of each node, $z\in C(y)$,
can be simply incremented without computing the intersection except for $\alpha_y$ itself. 
This is called {\bf descendant-intersect-skipping} (DIS).  
\item In Line~9, we continue the updating process.
In the recursive call, the intersection, $I$, is used instead of the original itemset, $E$. It follows that $\alpha_{y'}\cap E = \alpha_{y'}\cap I$ for 
each child $y'$ of $y$ since $I = \alpha_y \cap E$.
By reducing $E$ to $I$ (i.e., $I$ is a subset of $E$), 
the computational cost of the recursive call after Line~9 is reduced. This pruning technique is called {\bf masking}.
\item If $I = \emptyset$, every descendant of $y$ has no  items in common $E$ so the descendants need not be updated. This is called {\bf descendant-update-skipping} (DUS).
\item Line~11 checks if $E \subseteq \alpha_y$ or not. 
If so, every right sibling of $y$ need not be updated.
This follows from the observation in Proposition~\ref{inclusion_prop2} that the entry for any subset of $E$ to be updated never appears in the successors of $y$. 
This is called {\bf successor-update-skipping} (SUS). 
\item Finally, if there is no entry for $E$ in $W(i)$, the new entry for $E$ is added as the right-most child of $x$.
Note that if $x$ is the root node, we set $c_x = 0$ and $\Delta_x = \Delta(i)$.
\end{itemize}
\begin{example}\label{ex_traversal}
We explain how Algorithm~\ref{weep_tree_const} works using 
$${\mathcal S}^3_4 = \langle \{1,2,3,5\},\{1,2,4\},\{2,3,4\},\{1,2,5\}\rangle.$$
$W(3)$ corresponds to the left tree in Figure~\ref{newEx}.
The function $update(root, t_4, W(3))$ is called to derive $W(4)$ from $W(3)$.
For the left-most child $e_1$, the intersection, $I_1 = \{1, 2, 5\}$ of $e_1$ with $t_4$ is computed. 
Since $I_1 \neq \emptyset$, we call $update(e_1, I_1, W(3))$ as shown in Line~9.
For the left-most child, $e_2$, of $e_1$, the intersection $I_2 = \{1, 2\}$ of $e_2$ with $I_1$ is computed using $I_1$ by masking.
Since $I_2 = \alpha_{e_2}$, DIS is applied in Lines~5-7. 
Then, $c_{e_5}$ is simply incremented by one. 
Moving to the right sibling, $e_4$, the intersection $I_4 = \{2\}$ of $e_4$ with $I_2$ is computed.
Since $I_4 \neq \emptyset$, $update(e_4, I_4, W(3))$ is called. 
Since $e_4$ does not have any children, the algorithm checks if there exists the entry for $I_4$ then backtrack to the second call (i.e., $update(e_1, I_1, W(3))$).  
Since there is no sibling of $e_4$, the algorithm returns to Line~15 and a new entry, $e_7$, for $I_1$ is added as the right most child of $e_1$. 
After backtracking to the first call, $SUS$ is applied in Line~11 since $I_1 = t_4$.
Thus, the updating of the two right-most sibling nodes, $e_3$ and $e_6$, is skipped and the algorithm proceeds to Line~15. 
Since an entry for $t_4$ exists, the updated tree, $W(4)$, is returned as the output. 
 
\begin{figure}[h]
		\begin{center}
		 	\vspace{-15mm}
	  		\includegraphics[width=9.5cm]{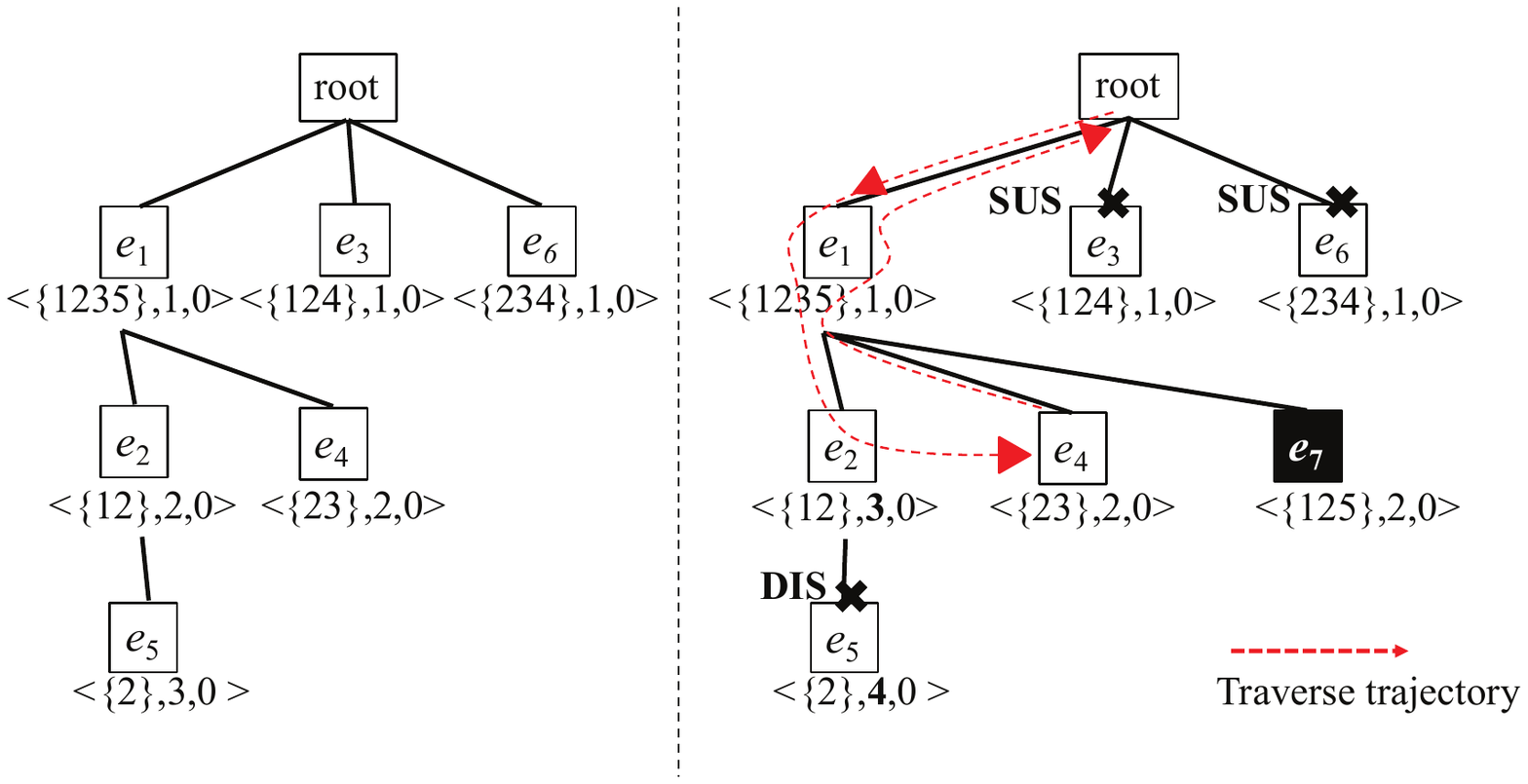}
 			\vspace{-15mm}
			\caption{The weeping trees $W(3)$ and $W(4)$ wrt ${\mathcal S}^3_4$}
			\label{newEx}
		\end{center}
\vspace{-5mm}
\end{figure}	
\end{example}
In this way, the $update$ function realizes the incremental intersection. 
Note that every node in $W(i)$ is visited at most once, which implies that Algorithm~\ref{weep_tree_const} efficiently runs 
$update(root, t_{i+1}, W(i))$ to return $W(i+1)$ in $O(kL)$ time.

\begin{figure}[h]
	\begin{center}
		\includegraphics[width=8.5cm]{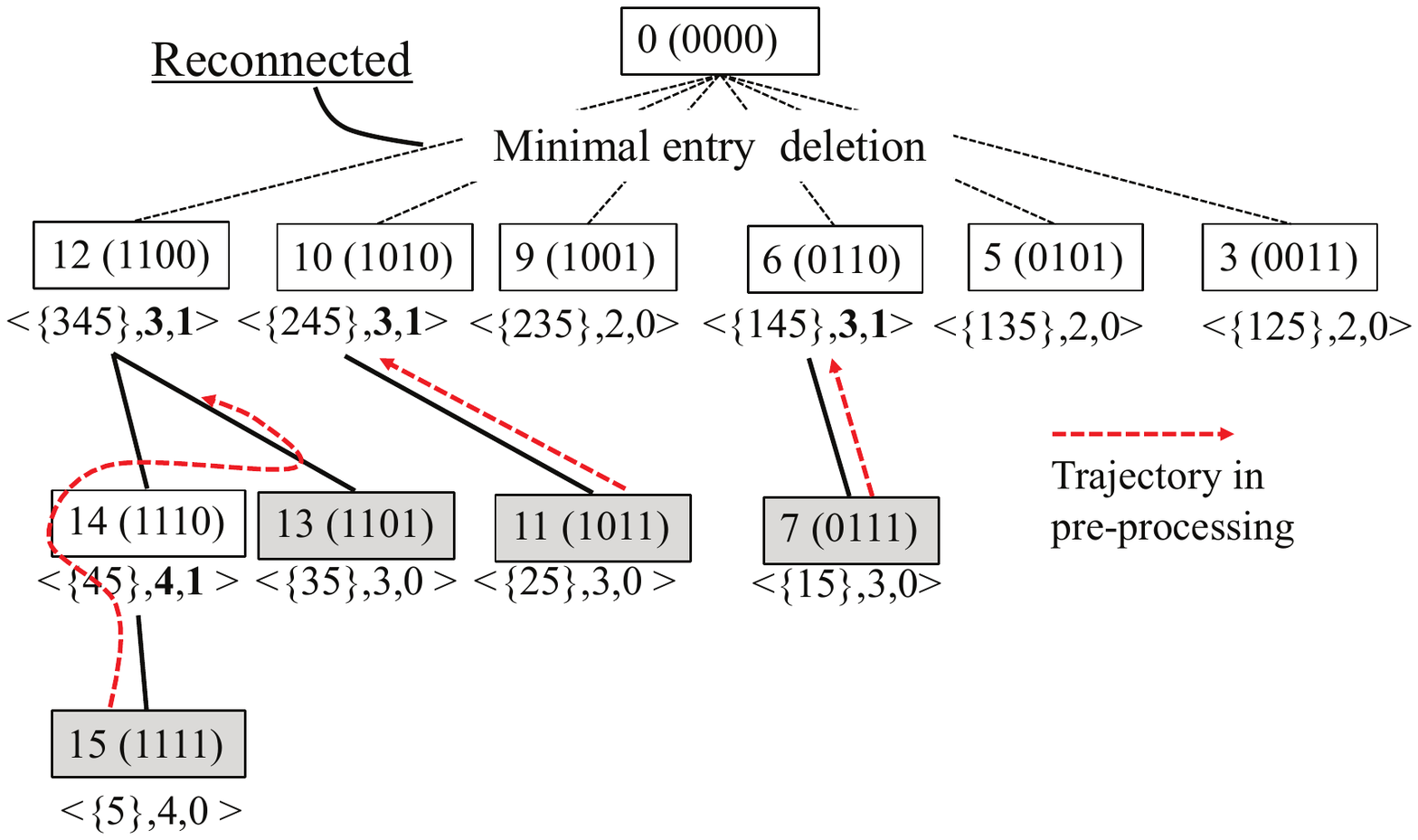}
						\vspace{-5mm}
		\caption{The reduced weeping tree of $W(4)$}
		
      	\label{red_weeping_prefix_fig}
	\end{center}
\end{figure}

Next, we show how PARASOL realizes the minimum entry deletion in the weeping tree.
As explained before, the minimum entries to be deleted are allocated in the shallowest layer relative to the root. 
Recall Example~\ref{weep_prefix_ex}, in which $\epsilon$ was set to 0.25 and PARASOL was used to delete the entries
with frequency counts of one at time $i = 4$. 
These minimum entries can be quickly accessed by applying min-heap to the shallowest layer. 
The reduced weeping tree is obtained by reconnecting the children of the deleted nodes with the root  as shown in Figure~\ref{red_weeping_prefix_fig}). 

Finally, the weeping tree is used as a pre-processing step to prepare for the $\Delta$-compression (i.e., 
seeking the tree for such a node that is $\Delta(n)$-covered by the parent).
Figure~\ref{red_weeping_prefix_fig} shows the reduced tree obtained by traversing the weeping tree $W(4)$
in a bottom up manner from the left-most leaf to the root; this process results in the removal of four nodes ($7$, $11$, $13$, and $15$).  
Note that $\Delta$-compression requires $O(k^2)$ time to completely check every pair of nodes. 
Thus, it is reasonable to carry out the $\Delta$-compression in a two-step procedure; first checking the parent-child $\Delta(n)$-covering to remove the $\Delta(n)$-covered children in a one-time traversal and subsequently performing the brute-force search of the remaining nodes.  


\section{Experiment}
Now, we empirically evaluate the performances of the baseline algorithm, PARASOL, and $\Delta$-compression, respectively.
They have been implemented with the weeping tree (in C language\footnote{available from https://github.com/yoshi-3/vldb2019.git}).
Ten real datasets were collected from \cite{yamamoto14} (Nos.~1-2), the FIMI repository \cite{fimi} (Nos.~3-9), and a new 
benchmark (No.~10) created from the Yahoo! Hadoop grid logs dataset \cite{yahoo} as shown in Table~\ref{static_table}. 
Note that $n$, $|I|$, $L$, and $L_{ave}$ are the end time, the number of different types of items, the maximum transaction length, and the average transaction length, respectively.   
\begin{table}[h]
	\begin{center}
	\begin{tabular}{|c|c|c|c|c|c|}
	\hline
   No & Datasets & $n$ & $|I|$ & $L$ & $L_{ave}$  \\
	\hline\hline
	1 & earthquake & 16,764   & 1,228   & 73     & 2.5   \\
	2 & weblog       & 19,465   & 9,958   & 106    & 18.2  \\
	3 & retail     & 88,162   & 16,469  & 76     & 10.3  \\
	4 & accidents	 & 340,183	 & 468	  & 51	   & 33.8	\\
    5 & chess	    & 3,196	    & 75	     & 37	   & 37		\\
    6 & connect	    & 67,557	 & 129	  & 43	   & 43    \\
    7 & kosarak	    & 990,002	 & 41,270 & 2,498	& 8.1   \\
    8 & pumsb	    & 49,046	 & 2,112  & 74	   & 73.6  \\
    9 & mushroom	 & 8,124    &	119	  & 23	   & 23    \\
    10 & hadoop      & 3,655,760 & 1,941,291 & 1,000 & 30.6\\

	\hline\hline 
	\end{tabular}
	\end{center}
		\caption{Characteristics of the used datasets}
			\label{static_table}     		
\end{table}

First, the baseline algorithm was compared with the state-of-the-art RC method, Skip LC-SS \cite{yamamoto14}, on the existing real datasets (Nos. 1-9).
Then, the scalability of the baseline algorithm was demonstrated on the synthesized datasets generated by the IBM market basket generator.
Next, the performance of PARASOL was compared with those of the baseline algorithm and the state-of-the-art PC method, MOA-IncMine~\cite{quadrana15}, when applied to the large-scale dataset (No.~10). Finally, we evaluated the effectiveness of $\Delta$-compression in reducing the output.

Table~\ref{baseline_comparison} describes the overall improvement that has been realized by the baseline algorithm, compared with Skip LC-SS, for each dataset in terms of the maximum error, $\Delta(n)$, the error ratio, $\frac{\Delta(n)}{n}$, and the average updating time (msec). 
In this experiment, the size constant, $k$, was fixed as 12,000. The results show that the baseline algorithm  drastically reducd the error.
Notably, in datasets 4, 5, 6, 8, and 9, the error ratio with Skip LC-SS reached one, meaning that the FIs could not be found for any $\sigma$ ($0 < \sigma < 1$), while the baseline provided a $\Delta(n)$-covered set of the FIs for which $\Delta(n)\leq\sigma n$. 
Moreover, the updating time of the baseline algorithm was lower than the one of Skip LC-SS.

\begin{figure*}[t]
\centering
\subfigure[Skip LC-SS versus the baseline]{
	  	\includegraphics[width=5.5cm]{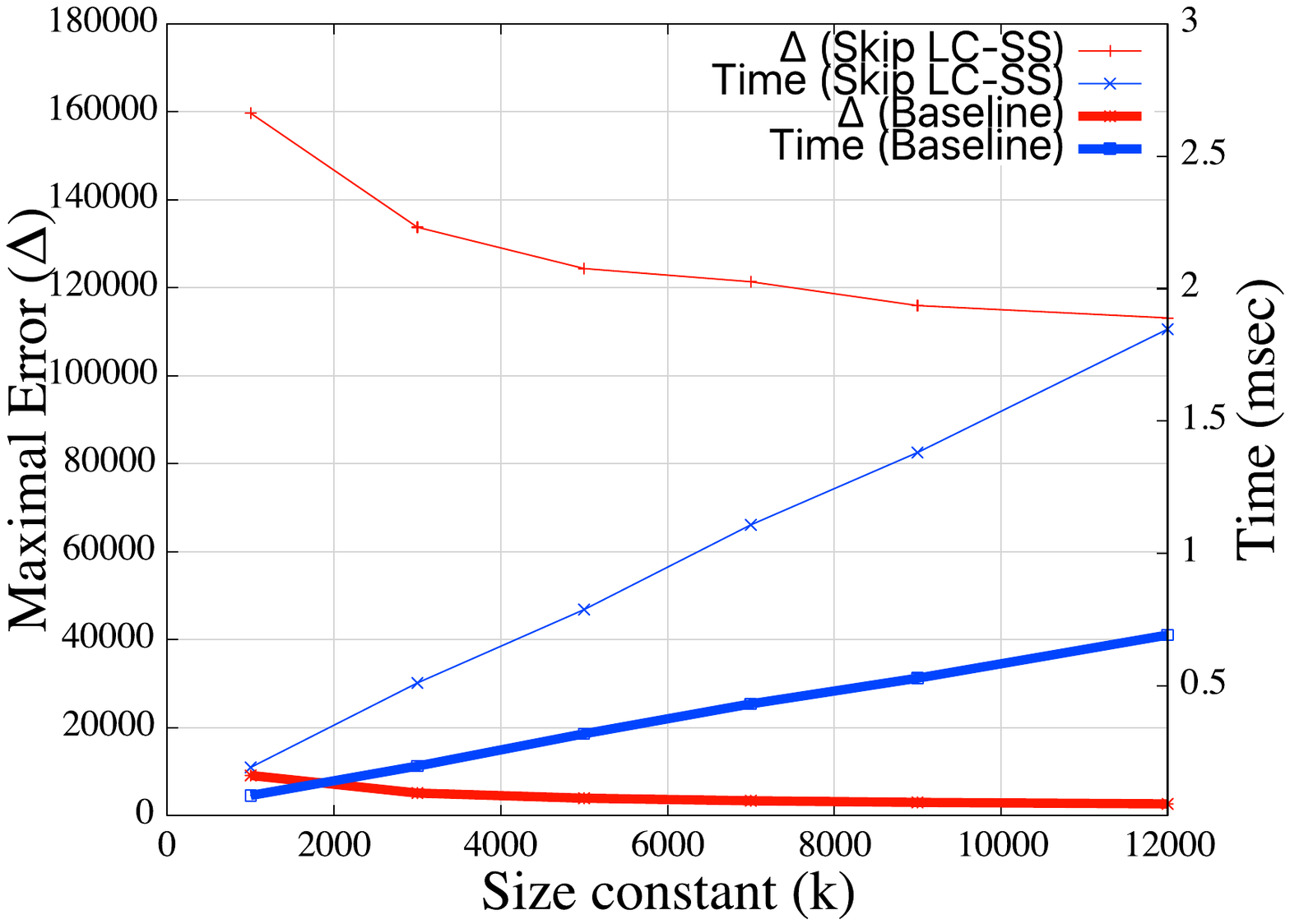}
				\label{k_7kosarak_ro}
}
\subfigure[Scalability with respect to $L$]{
	  	\includegraphics[width=5.5cm]{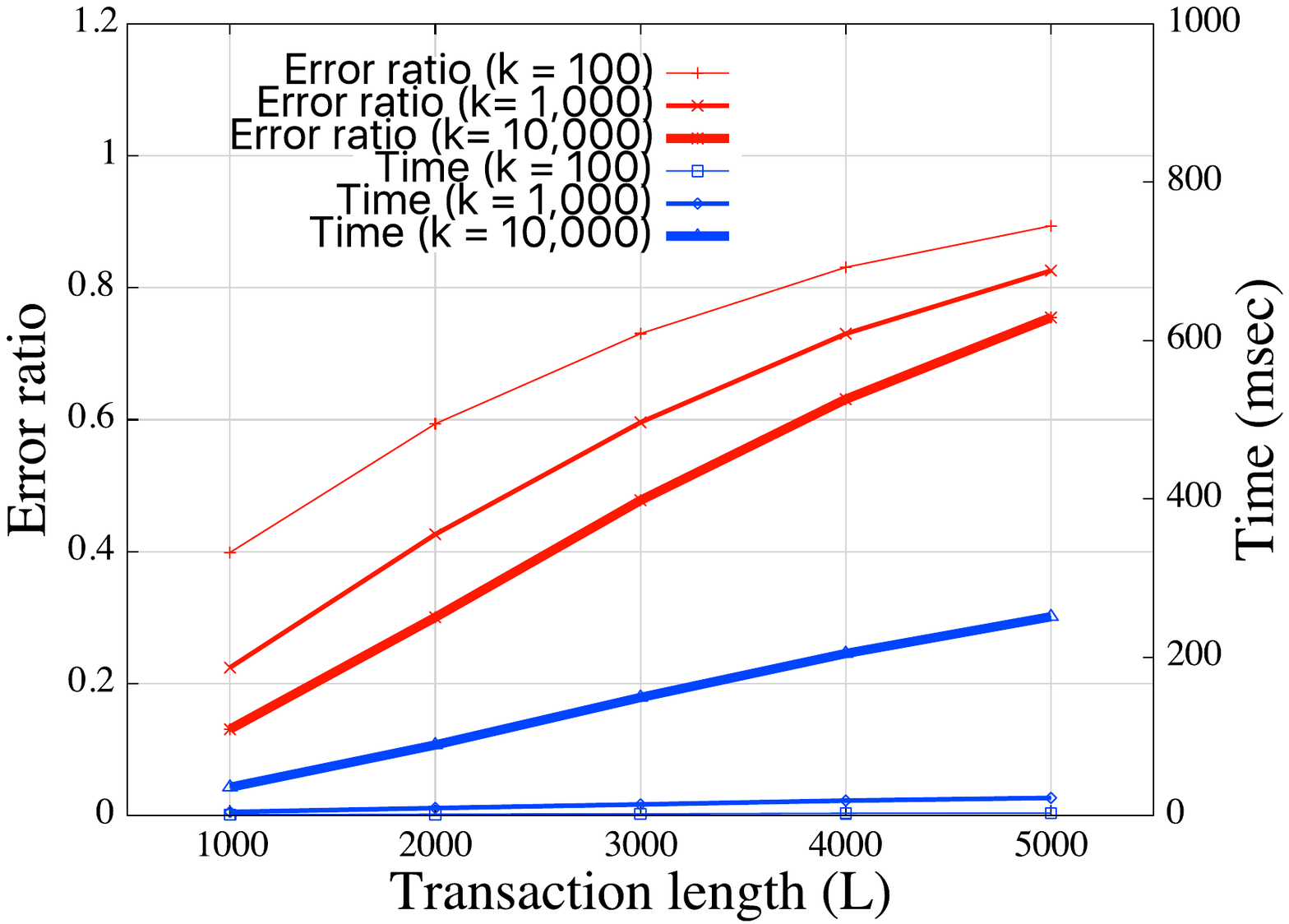}
				\label{k_ro}
}
\subfigure[Scalability with respect to $k$]{
	  	\includegraphics[width=5.5cm]{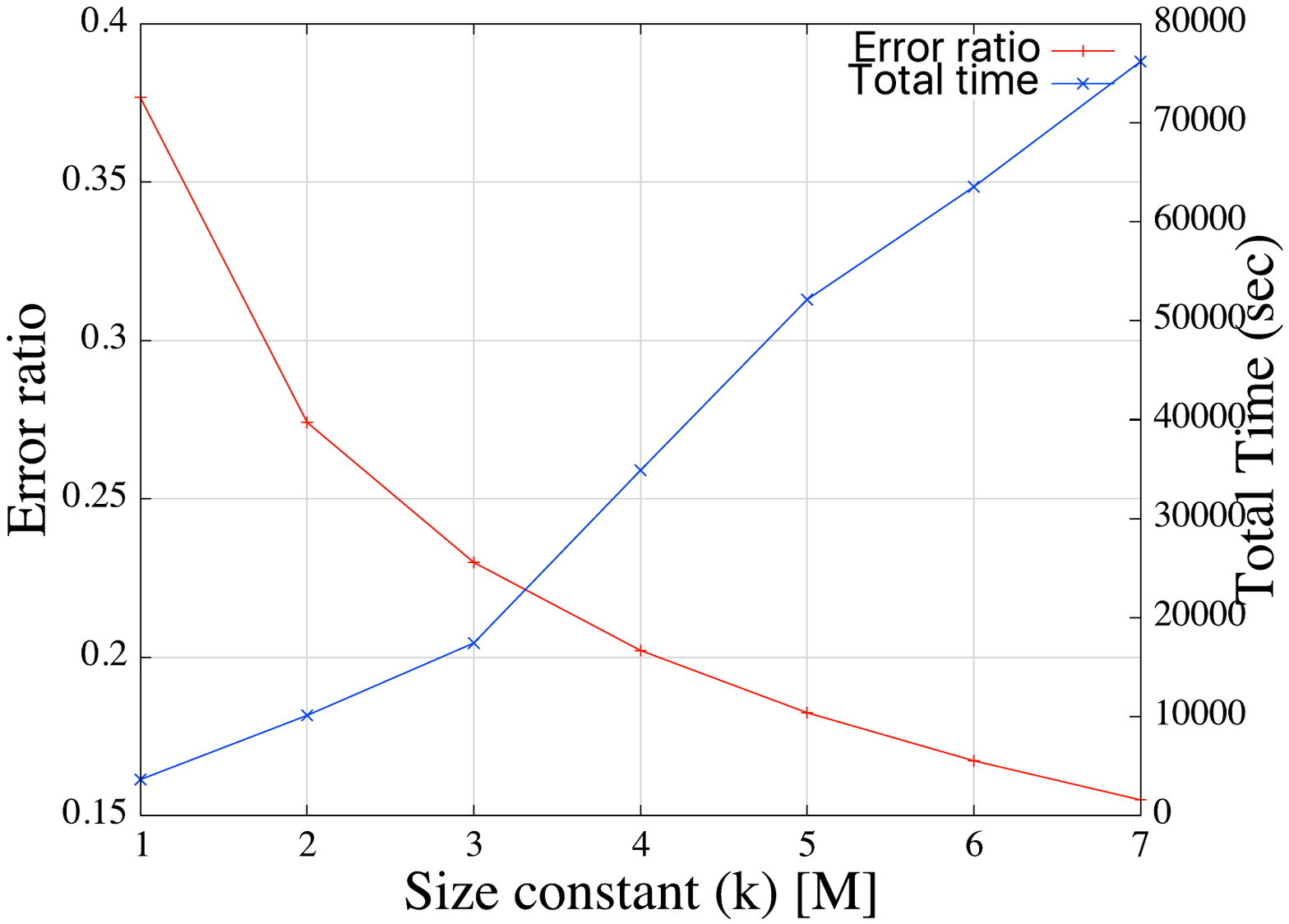}
				\label{k_6connect_ro}
}
			\vspace{-2mm}
\caption{Performance of the baseline algorithm}
\end{figure*}

\begin{table}
	\begin{center}
	\begin{tabular}{|c|c|c|c|c|c|c|}
	\hline
	  & \multicolumn{3}{|c|}{Skip LC-SS} & \multicolumn{3}{|c|}{Baseline algorithm}  \\
   	\hline
 	No  & $\Delta(n)$ & $\Delta(n)$/$n$ & Time & $\Delta(n)$ & $\Delta(n)$/$n$ & Time \\
 \hline\hline
1  & 157 & 0.0094 & 0.15 & 20 & 0.0012 & 0.23 \\
2  & 11,792 & 0.61 & 10 & 2,953 & 0.15 & 0.29\\
3  & 23,987 & 0.27 & 3.8 & 156 & 0.0018 & 1.20\\
4  & 340,182 & {\bf 1} & 16 & 179,683 & 0.53 & 1.40\\
5  & 3,196 & {\bf 1} & 20 & 2,434 & 0.76 & 4.50 \\
6  & 67,556 & {\bf 1} & 18 & 62,736 & 0.93 & 2.10 \\
7  & 113,143 & 0.11 & 1.8 & 2,681 & 0.0027 & 0.69\\
8  & 49,046 & {\bf 1} & 18 & 42,513 & 0.87 & 2.70\\
9  & 8,124 & {\bf 1} & 15 & 492 & 0.061 & 1.80\\
\hline\hline
	\end{tabular}
	\end{center}			
	\caption{Skip LC-SS versus The baseline algorithm}
	\label{baseline_comparison}
\end{table}
Skip LC-SS was compared with the baseline algorithm with a range of size constants for the dataset 7, on which Skip LC-SS performed the best in terms of the error ratio. The results are summarized in Figure~\ref{k_7kosarak_ro}.
The data shows that the baseline algorithm (bold lines) performed better than Skip LC-SS in terms of maximum error and execution time for all values of $k$. Moreover, the execution time decreased linearly and $\Delta(n)$ increased as $k$ was decreased.

Next, the scalability of the baseline algorithm was evaluated by varying the maximum transaction length, $L$, as it is applied to the synthesis data. This dataset was generated by the IBM market basket generator; its stream size, $n$, was 10,000 and the number of types of items, $|I|$, was about 24,000. Figure~\ref{k_ro} describes the average updating time and the error ratio, $\frac{\Delta(n)}{n}$, as functions of $L$. As the time complexity for the updating process is $O(kL)$, the execution time linearly increased as $L$ was increased. In an ordinary problem setting, we need to focus the solution space on $2^L$ itemsets for $L$, resulting in a rapid increase in the error count. In contrast, the proposed solution can tolerate such large transactions that always contain 5,000 items.

In addition, the change in the error ratio as a function of $k$ in the range of 1,000,000 to 7,000,000 was evaluated with the dataset 6, for which the error ratio achieved by the baseline algorithm was the largest. The results are shown in Figure~\ref{k_6connect_ro}.
Then, the error ratio linearly decreased while the execution time increased as k was increased. Note that a simple open MP parallelization was used for this computation. 
Such a parallelization scheme can be used in RC approximation with a fixed memory resource~\cite{yamamoto16} but the issue of efficient implementation of such a scheme is beyond the scope of this paper.

\begin{figure}[t]
	\begin{center}
	 	\vspace{-30mm}
	  	\includegraphics[width=7.5cm]{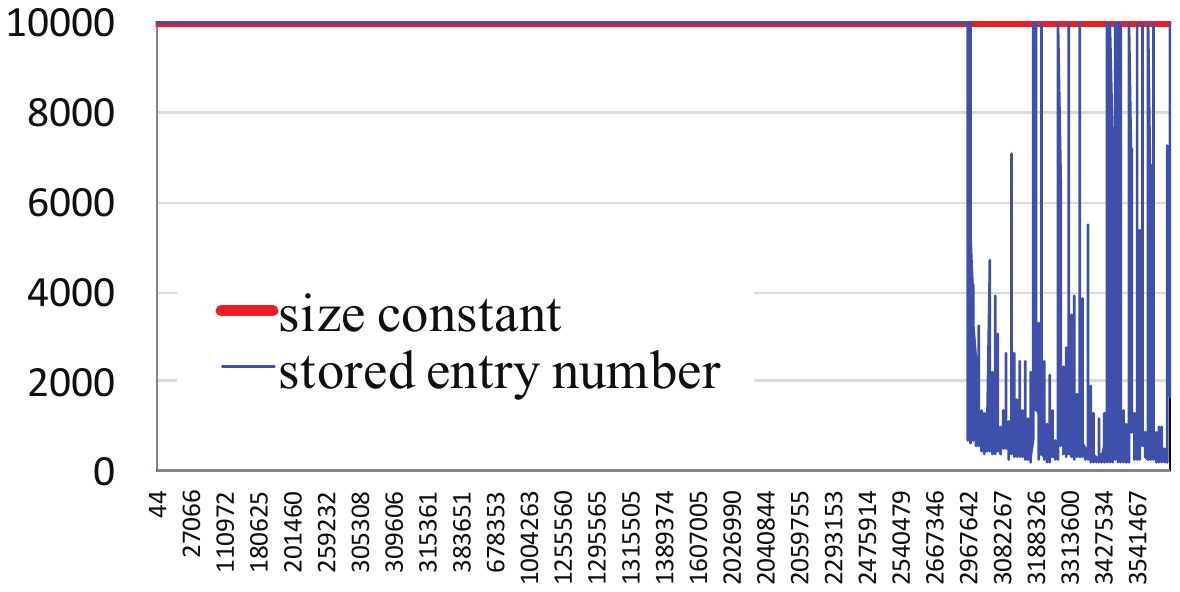}
		 		\vspace{-30mm}
			\caption{Time series of the stored entries, $k$}		
		\label{entrysize}
	\end{center}
		\begin{center}
		 \vspace{-30mm}
	  	\includegraphics[width=7.5cm]{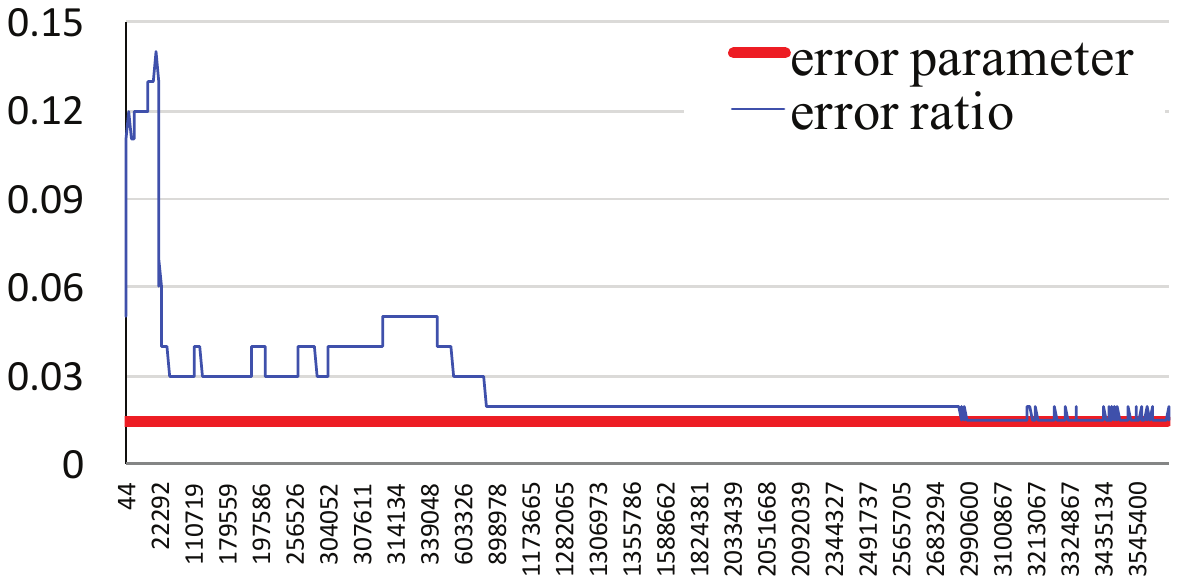}
		 		\vspace{-30mm}
			\caption{Time series of the error ratio, $\Delta$}
		\label{errorcount}
	\end{center}
	 \vspace{-5mm}
\end{figure}

Next, we compare PARASOL with the baseline algorithm and MOA-IncMine using the hadoop dataset (No.~10).
Figures~\ref{entrysize} and~\ref{errorcount} show the time series changes of the number of used entries, $k(i)$, 
and the error ratio, $\frac{\Delta(i)}{i}$, respectively, while $k = 10,000$ and $\epsilon = 0.015$. 
The results show that PARASOL violates the set $\epsilon$ value due to the RC approximation that was applied earlier in the time series period but, after this period, the error ratio recovers to $\epsilon$, this autonomous recover of $\epsilon$ is characteristic of PARASOL. Moreover, $k(i)$ became smaller than $k$ due to the PC approximation in PARASOL.
Note that PARASOL composes $T(n)$ of 1,626 entries whike the baseline algorithm always uses the 10,000 entries.
In contrast, MOA-IncMine generates 770,995 SFCIs for some intermediate segment when $\epsilon = 0.1$ and a segment size of $w = 1,000$; thus,  
it could not finish updating the SFCIs (this was true for other values of $w$, as well). 
The drastic decrease in performance due to streaming transactions (or segments) is characteristic of existing PC methods. PARASOL avoids this drawback and remains sustainable by switching between PC and RC approximations.

\begin{table}	\begin{center}
	\begin{tabular}{|c|c|c|c|c|c|}
	\hline
	  & \multicolumn{2}{|c|}{MOA-IncMine} & \multicolumn{3}{|c|}{PARASOL with $\Delta$-compression} \\
   	\hline
 	No  & Time & \#Num & $\Delta(n)/n$ & Time & \#Num ($k(n)$)\\ \hline\hline
	1  & 0.16 s & 19    & 0.1 & 1.7 s & 20 (23)\\
	2  & 2 m, 46 s & 16,270 & 0.1 & 12.92 s & 182 (42,204)\\
	3  & 0.68 s & 58    & 0.1 &  3.02 s & 30 (50)\\
	4  & OM & -       & 0.374 & 114 m, 7 s & 4,194 (100k)\\
	5  & OM & -       & 0.611 & 9 m, 25 s & 2,780 (100k)\\
	6  & TO & -       & 0.761 & 44 m, 32 s & 3,833 (100k)\\
	7  & 5.56 s  & 14   & 0.1   & 29.21 s & 10 (15)\\
	8  & OM & -       & 0.779 & 29 m, 43 s & 7,437 (100k)\\
	9  & 2 m, 47 s & 3,430   & 0.1 & 8.49 s & 938 (6,290)\\
\hline\hline
	\end{tabular}
	\end{center}		
				\vspace{-2mm}	
	\caption{MOA-IncMine versus PARASOL}
	\label{parasol_comparison}
				\vspace{-2mm}
\end{table}

Table~\ref{parasol_comparison} compares the performance of MOA-IncMine with that of PARASOL with $\Delta$-compression when applied to the other datasets in terms of execution time and number of entries in the output. Given $\epsilon = 0.1$, MOA-IncMine was run with a 1 GB Java heap size for six hours (the segment size was fixed as the default (1,000)). PARASOL was run with $k = 100, 000$ and $\epsilon=0.1$.

The results show that MOA-IncMine results in out-of-memory (OM) and time-out (TO) exceptions when applied to datasets 4, 5, 6, and 8. On the other hand, PARASOL successfully processes all of these datasets although the error ratio, that MOA-IncMine causes the out-of-memory (abbreviated by OM) and time-out (abbreviated by TO) exceptions 
in No. 4, 5, 8 and No.~6 datasets.
PARASOL processes all, however the error ratio ($\frac{\Delta(n)}{n}$), exceeded the initial value of $\epsilon$. 
In datasets, 1, 3,~7, MOA-IncMine was faster than PARASOL; this is because these datasets were relatively sparse, implying that PARASOL tends to be more 
powerful for dense datasets, although the speed also depends on the segment size and size constant. 
The output of PARASOL was generally smaller than that of MOA-IncMine due to $\Delta$-compression as shown by number of used entries, $k(n)$.  

Table~\ref{table_pruning_effect} shows the pruning effect of using the weeping tree structure while k = 12,000 and $\epsilon = 0$.
The average ratio between traversing nodes and total nodes for each time and the average ratio between nodes that have non-empty intersections and total nodes for each time are shown. The results show that the use of the weeping tree structure is superior for pruning the search space in dense datasets (Nos. 2, 4, 5, 6, and 8) and does not impose any additional maintenance cost compared with the standard indexing technique.
The \#Num$^{-}$ and \#Num$^{+}$ columns show the number of entries after pre-processing and number of entries that were output by the $\Delta$-compression, respectively. For example in dataset 6, the ordinal 12,000 entries were reduced to 3,418 by pre-processing and 535 of these entries remained in the output after the second filtering step.
The Time$^{-}$ and Time$^{+}$ columns show the execution time of the $\Delta$-compression with and without the pre-processing step, respectively. 
\begin{table}[h]
	\begin{center}
	\label{table}
	\begin{tabular}{|c|c|c|c|c|c|c|}
	\hline
 	  &  Visiting    & Non- &  &   &  & \\
	 No   &   ratio  & empty &   \#Num$^{-}$  &  \#Num$^{+}$ &    Time$^{-} $           & Time$^{+} $\\ 
	    &    (\%)      & ratio   &          &              &      (sec)              & (sec)  \\
    \hline\hline
1 & 38.4 &5.0  & 5,129 &3,895     &2.23  & 0.89 \\
2 & 4.8  &13.3  & 345  &138      &1.05  &0.00 \\
3 & 68.6 &22.6 & 9,367 &6,503     &4.39 & 3.36 \\
4 & 19.9 &52.0 & 2,010 &626       &1.49 & 0.07\\ 
5 & 41.8 &55.1 & 1,468 &446       &1.63 & 0.05 \\
6 & 42.9 &62.9 & 3,418 &535      &1.48 & 0.16 \\
7 & 32.6 &15.0 & 5,589 &3,258     &2.17 & 0.74 \\
8 & 44.4 &57.9 & 6,004 &1,267     &2.26 & 0.7\\
9 & 36.4 &12.5 & 4,417 &2,251     &3.33 & 0.75\\ 
\hline\hline

	\end{tabular}
	\end{center}
			\caption{Pruning effects of using the weeping tree}
			\label{table_pruning_effect}
\end{table}



\section{Concluding remarks}
In this study, we have proposed a novel solution for FIM-SD that involves seeking a $\Delta$-deficient condensed representation, from which every FI and its frequency can be restored, while bounding the maximum error by an integer, $\Delta$. While the existing FIM-SD methods are limited to strictly PC and RC methods, we have mitigated the drawbacks of each approach by introducing a unified PC and RC-approximation scheme called PARASOL. We furthermore introduce a post-filtering technique called $\Delta$-compression and a novel data structure called the weeping tree. Experimental trials on ten datasets show that the proposed technique outperforms the existing FIM-SD methods. Moreover, the proposed algorithm is scalable in terms of the transactional length; in other words, it can tolerate any bursty transaction without running into an OM exception. This feature is favorable for analyzing large volumes of streaming transactions that consistently contain many items.
Such large streaming transactions are often encountered in surveillance domains of sensor networks and cloud servers. In addition, along with the recent success achieved with deep learning, it is now necessary to analyze the cognitive correlations among mid-level objects for explanatory domains. 
We believe that integrating modern online learning with the solution presented here will give rise to a new methodology for streaming data analysis. 
From a technical standpoint, it will be important to introduce novel parallel-processing techniques to further improve the scalability of this technique. In addition, it will be fruitful to study how such techniques can be efficiently embedded into the proposed weeping tree structure.
\section*{Appendix}

The $\Delta$-covered set has an interesting relationship with the so-called {\it $\Delta$-closed set}~\cite{boley09}, which is known as a condensed representation for FIs. 
An itemset $\alpha$ is {\it $\Delta$-closed} if 
there exists no itemset $\beta$ such that $\beta\succeq_{\Delta}^n\alpha$. 
Let ${\cal F}_{\Delta}$ denote the family of $\Delta$-closed FIs.
${\cal F}_{\Delta}$ is uniquely determined but does not have the $\Delta$-deficiency property.
Consider again Example~\ref{ex2} and assume that every $\alpha_i\in P$ is frequent by setting $\sigma$ as 0.5. 
Thus, we have ${\cal F}_1 = \{\alpha_4\}$, since $\alpha_i$ ($1\leq i \leq 3$) is not 1-closed.
Accordingly, ${\cal F}_1$ is no longer a $1$-covered set of ${\cal F}$, since the frequency of $\alpha_1$ cannot be 
restored from $\alpha_4$ within the error range below one.  

\begin{proposition}\label{prop_closed_set}
Let $\cal F$ be the family of FIs and $\Delta$ an integer. 
Then, ${\cal F}_{\Delta}$ is included by any $\Delta$-covered set of ${\cal F}$.
\end{proposition}
{\it Proof.} 
Assume that there exist a $\Delta$-closed FI $\alpha$ such that $\alpha\not\in T$ for some $\Delta$-covered set 
$T$ of ${\cal F}$. Since $\alpha$ is a FI but is not included in $T$, there should exist a $\beta\in T$ such that 
$\beta\succeq_{\Delta}^n \alpha$. However, this contradicts that $\alpha$ is $\Delta$-closed. $\Box$


\end{document}